\documentclass[%
reprint,
superscriptaddress,
 amsmath,amssymb,
 aps,
]{revtex4-2}
\usepackage{graphicx}
\usepackage{dcolumn}
\usepackage{bm}
\usepackage{subfigure}

\RequirePackage[normalem]{ulem}
\RequirePackage{color}
\newcommand{\new}[1]{{\textcolor{blue}{#1}}}

\usepackage[dvipsnames]{xcolor}
\usepackage{xcolor}

\begin{document}

\renewcommand{\vec}{\mathbf}

\title{Observation of replica symmetry breaking in standard mode-locked fiber laser}



\author{Nicolas P. Alves}
\affiliation{%
 Departamento de F\'{\i}sica, Universidade Federal de Pernambuco, 50670-901 Recife, PE, Brazil
}%

\author{Wesley F. Alves}
\affiliation{%
 Departamento de F\'{\i}sica, Universidade Federal de Pernambuco, 50670-901 Recife, PE, Brazil
}%

\author{Andr\'{e} C. A. Siqueira}
\affiliation{%
 Departamento de F\'{\i}sica, Universidade Federal de Pernambuco, 50670-901 Recife, PE, Brazil
}%

\author{Naudson L. L. Matias}
\affiliation{%
 Departamento de F\'{\i}sica, Universidade Federal de Pernambuco, 50670-901 Recife, PE, Brazil
}%

\author{Anderson S. L. Gomes}
\affiliation{%
 Departamento de F\'{\i}sica, Universidade Federal de Pernambuco, 50670-901 Recife, PE, Brazil
}%

\author{Ernesto P. Raposo}
\affiliation{Laborat\'orio de F\'{\i}sica Te\'orica e Computacional, Departamento de F\'{\i}sica, Universidade Federal de Pernambuco, 50670-901 Recife, PE, Brazil}

\author{Marcio H. G. de Miranda}
\affiliation{%
 Departamento de F\'{\i}sica, Universidade Federal de Pernambuco, 50670-901 Recife, PE, Brazil
}%
\email{marcio.miranda@ufpe.br}

\date{February 29, 2024}

\begin{abstract}
We report the first experimental demonstration of the replica symmetry breaking (RSB) phenomenon in a fiber laser system supporting standard mode-locking (SML) regime.
Though theoretically predicted, this photonic glassy phase remained experimentally undisclosed so far. 
We employ an ytterbium-based mode-locked fiber laser with a very rich phase diagram. 
Two phase transitions are observed separating three different regimes: CW, quasi-mode-locking (QML), and~SML. 
The regimes are intrinsically related to the distinct dynamics of intensity fluctuations in the laser spectra. 
We set the connection between the RSB glassy phase with frustrated modes and onset of L-shaped intensity distributions in the QML regime, which impact directly the replica overlap measure. 
\end{abstract}

\maketitle



The theoretical proposal~\cite{a61,a63} and experimental demonstration~\cite{a66,a68} of the  replica symmetry breaking (RSB) phenomenon in the photonic context have made a great impact in the field of complex systems~\cite{a81c,a59,a1c,a1,a1b}.
In Parisi's approach to disordered magnetic spin glasses~\cite{a80}, RSB occurs when replicas of the system prepared under identical conditions can reach different states in a free energy landscape with multiple local minima that can trap a spin configuration for long times. 
In the magnetic-to-photonic analogy employing random lasers (RLs), the amplitude of the optical modes and input excitation energy play,  respectively, the roles of the spins and inverse temperature~\cite{a61,a63}. 
Remarkably, as~commented by Parisi~\cite{a81c}, experimental evidence of RSB was provided by multimode RLs, since in this case it is possible to observe the modes occupancy and thus measure directly the order parameter function with RSB signature.

The photonic analogue of the magnetic RSB spin glass phase, with correlated random spins prevented to align parallel, corresponds to the RSB glassy behavior in~RLs, in which the synchronous oscillation of nontrivially correlated modes is frustrated~\cite{a61,a63}. 
In contrast, the passive standard mode-locking (SML) lasing regime~\cite{ml,gor}, with high phase coherence of modes locked in phase, is analogous to the replica-symmetric ferromagnetic phase displaying parallel spins. 
In this sense, the concept of magnetic or photonic frustration relates to configuration states in which the full alignment of spins or synchronous oscillation of modes is hampered.

As the interest in photonic systems with disorder and nonlinearity has increased substantially in recent years~\cite{a1,a1b}, different types of materials have been investigated, such as cavityless RLs, e.g., 2D amorphous solid-state T5OCx oligomer~\cite{a66,a1c} and 3D YBO compound doped with crystalline powder of Nd$^{3+}$ ions~\cite{a68,bbbb}, as well as the closed-cavity multimode Q-switched Nd:YAG laser~\cite{a71}. 
An erbium-doped random fiber laser has been also exploited to demonstrate RSB~\cite{anderson}, 
as well as a hybrid electronically addressable random fiber laser~\cite{edwin}. 
All these systems present a replica symmetric regime in the prelasing phase and RSB above threshold. 

In RL systems the RSB regime is glassy, with frustrated coherent oscillation of lasing modes displaying nontrivial correlations. 
In striking contrast, in the multimode Nd:YAG system the RSB regime is analogous to the random bond ferromagnet, with most spins pointing parallel.
In this case, the active modes present considerable degree of coherence, emission occurs in the form of ultrashort pulses, and RSB arises as the random activation of a given subset of coherent modes after each excitation pulse hampers the others~\cite{a71}.
We stress that the photonic behavior of the Q-switched Nd:YAG system~\cite{a71}, with absence of a saturable absorber in the closed cavity, does {\it not} correspond to the SML regime of locked modes with high phase coherence~\cite{ml,gor}. 

Here we consider an ytterbium-based mode-locked fiber laser (MLFL) supporting SML behavior, which presents both nonlinearity and disorder ingredients coming from the competition for gain and frustration among modes, thus opening the possibility for RSB. 
Indeed, to our knowledge, there is no experimental study so far on the manifestation of RSB in SML lasers, although its existence has been predicted theoretically~\cite{a63,a59,a1c}. 
To make explicit the contrast with the current experimental scenario, no RL system investigated to date has been shown to support both SML and RSB glassy behaviors~\cite{a1,a1b}. 
Moreover, in our Yb-based SML system the RSB regime is spin-glass-like with frustrated modes, while in the Q-\linebreak switched Nd:YAG laser~\cite{a71} the RSB~regime is ferromag\-netic-like with considerable degree of phase coherence.

The Yb-based MLFL is a rich, complex dynamic system, with regimes spanning from deterministic chaos to multi-pulsing behavior~\cite{Lucas18, li2010geometrical}. 
In this work, we report the first experimental evidence of RSB in a SML laser. 
As the excitation energy increases, we observe two phase transitions separating three different regimes: CW, quasi-mode-locking (QML), and SML. 
Their magnetic analogues are, respectively, the paramagnetic, RSB glassy, and replica-symmetric ferromagnetic phases, in agreement with the theoretical phase diagram proposed in~\cite{a63}. 
These photonic regimes are intrinsically related to the distinct dynamics of intensity fluctuations in the laser spectra. 
Indeed, while the CW~and SML regimes present Gaussian-like distributions of intensities, the QML behavior displays L-shaped distributions that impact directly the replica overlap correlation measure. 
%



The experimental setup consists of a homemade Yb-based MLFL system~\cite{Lucas18, Cecilia19} 
that achieves passive mode locking via nonlinear polarization rotation~\cite{Ferman}. 
The MLFL produces optical spectra centered around 1025~nm with a repetition rate of 120~MHz in the SML regime. 
The pump source is a CW fiber-coupled diode laser emitting at 976~nm.
Data acquisition is accomplished by directing the output laser beam and recording its optical spectrum using an OSA with 0.24~nm resolution (much shorter than all spectral bandwidths analyzed).
RF spectrum and pulse train are also measured for regime operation recognition. 
We consider 3000 spectra (replicas) for each excitation current.
Each spectrum corresponds to an integration time of~9~ms, with interval of 130~ms between replicas (see Supplemental Material~\new{\cite{sm}} 
for further details and results for other integration~times).

\begin{figure}[t]
\includegraphics[width=0.5\textwidth]{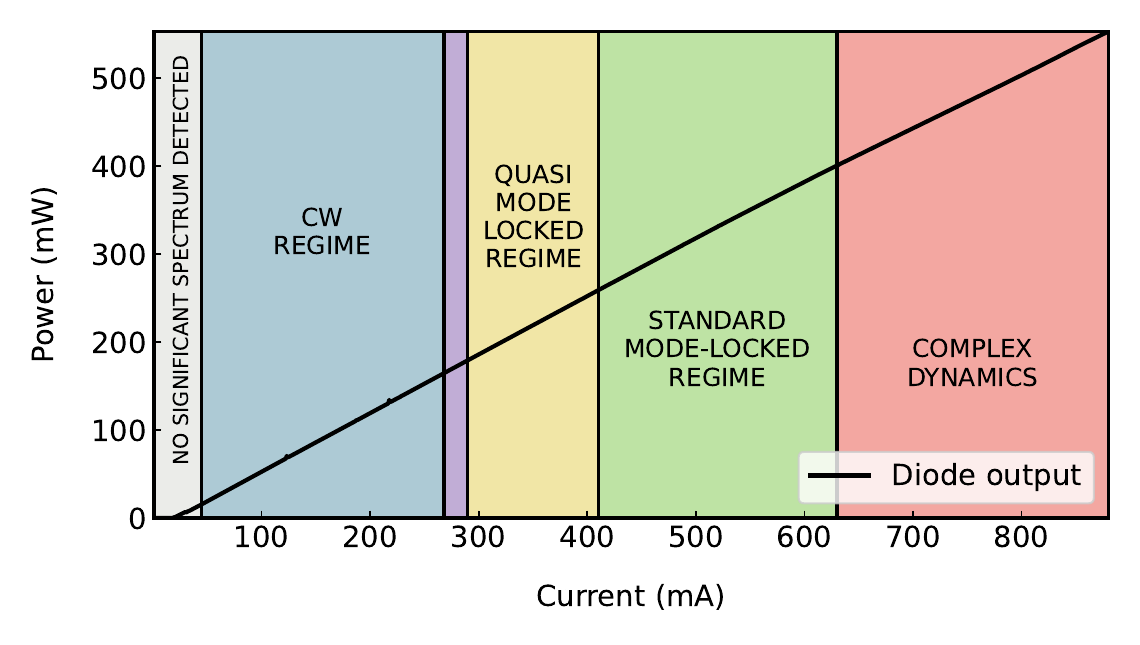}
\caption{Photonic regimes displayed by the Yb-based mode-locked fiber laser (MLFL) system.}
\label{f2}
\end{figure}

Figure~\ref{f2} shows the optical regimes displayed by the Yb-based MLFL as a function of the excitation current for the present choice of parameters (polarization,  dispersion). 
%
%
In particular, we note the CW, QML, and SML regimes, and also a complex dynamic phase for quite large currents that is out of the scope of this work. 
The purple region indicates a transition range in which the MLFL behaviour fluctuates between the CW and QML regimes.
The whole sets of 3000 spectra for the three relevant regimes are shown in Fig.~S2 of the Supplemental Material~\new{\cite{sm}}.
In particular, Figs.~S3 and S4 (temporal trace) indicate that the ultrashort pulses in the SML regime have an average duration of 275~fs.
In sharp contrast, the QML regime is not pulsed.
In fact, in the QML regime the multiple longitudinal modes take part~in a frustrated competition for gain to define the set of modes that will be locked in the SML phase at larger~currents.
%

The RSB phenomenon is investigated in the photonic context through the correlation measure between intensity fluctuations in Parisi's overlap parameter~\cite{a59},
\begin{equation} 
\label{eq:parisi}
q_{\gamma \beta} = \frac{\sum_{k} \Delta I_\gamma (\omega_k) \Delta I_\beta (\omega_k)}{ \sqrt{\sum_{k} [\Delta I_\gamma (\omega_k)]^2} \sqrt{ \sum_{k}[\Delta I_\beta (\omega_k)]^2} },
\end{equation}
where $\gamma, \beta = 1, 2, ..., N_{R}$ are the spectrum (replica) lables ($N_{R} = 3000$). 
The mean intensity for a given frequency~$\omega_k$ indexed by $k$ reads $\langle I_\gamma (\omega_k) \rangle =\sum_{\gamma = 1}^{N_{R}}I_{\gamma}(\omega_k)/N_{R}$, thus setting the intensity fluctuations $\Delta I_\gamma (\omega_k) = I_\gamma (\omega_k) - \langle I_\gamma (\omega_k) \rangle$. 
By considering all pairs of distinct replicas~$\gamma$ and~$\beta$, the distribution~$P(q)$ of $q_{\gamma \beta}$~values can be built. 
A single maximum of $P(q)$ at $q = 0$, or two~side maxima around $q = \pm 1$, are indicative of the replica symmetric or RSB regime, respectively. 

The RSB analysis can be complemented by the Pearson correlation coefficient~\cite{edwin},
\begin{equation}
C_{k_i k_j} = \frac{\sum_{\gamma} \Delta I_\gamma (\omega_{k_i}) \Delta I_\gamma (\omega_{k_j}) }{ { \sqrt{\sum_{\gamma} [\Delta I_\gamma (\omega_{k_i})]^2} \sqrt{ \sum_{\gamma}[\Delta I_\gamma (\omega_{k_j})]^2} } }.
\label{eq:pearson}
\end{equation}
in which the indexes $k_i$ and $k_j$ label, respectively, the frequencies $\omega_{k_i}$ and~$\omega_{k_j}$ in the {\it same replica}~$\gamma$. 
Albeit similar in form, we note that Eqs.~(\ref{eq:parisi}) and~(\ref{eq:pearson}) are subtly different: while in Parisi parameter sums are over frequencies, \linebreak in the Pearson coefficient sums span over spectra acquired at different times. 
A negligible $C_{k_i k_j}$ implies that fluctuations at distinct frequencies behave nearly uncorrelated. 
On the other hand, a positive (negative)~$C_{k_i k_j}$ indicates that fluctuations at a given frequency are related to fluctuations with same (different) sign at a distinct frequency of the same spectrum.
%

\begin{figure}[t]
\includegraphics[width=0.5\textwidth]{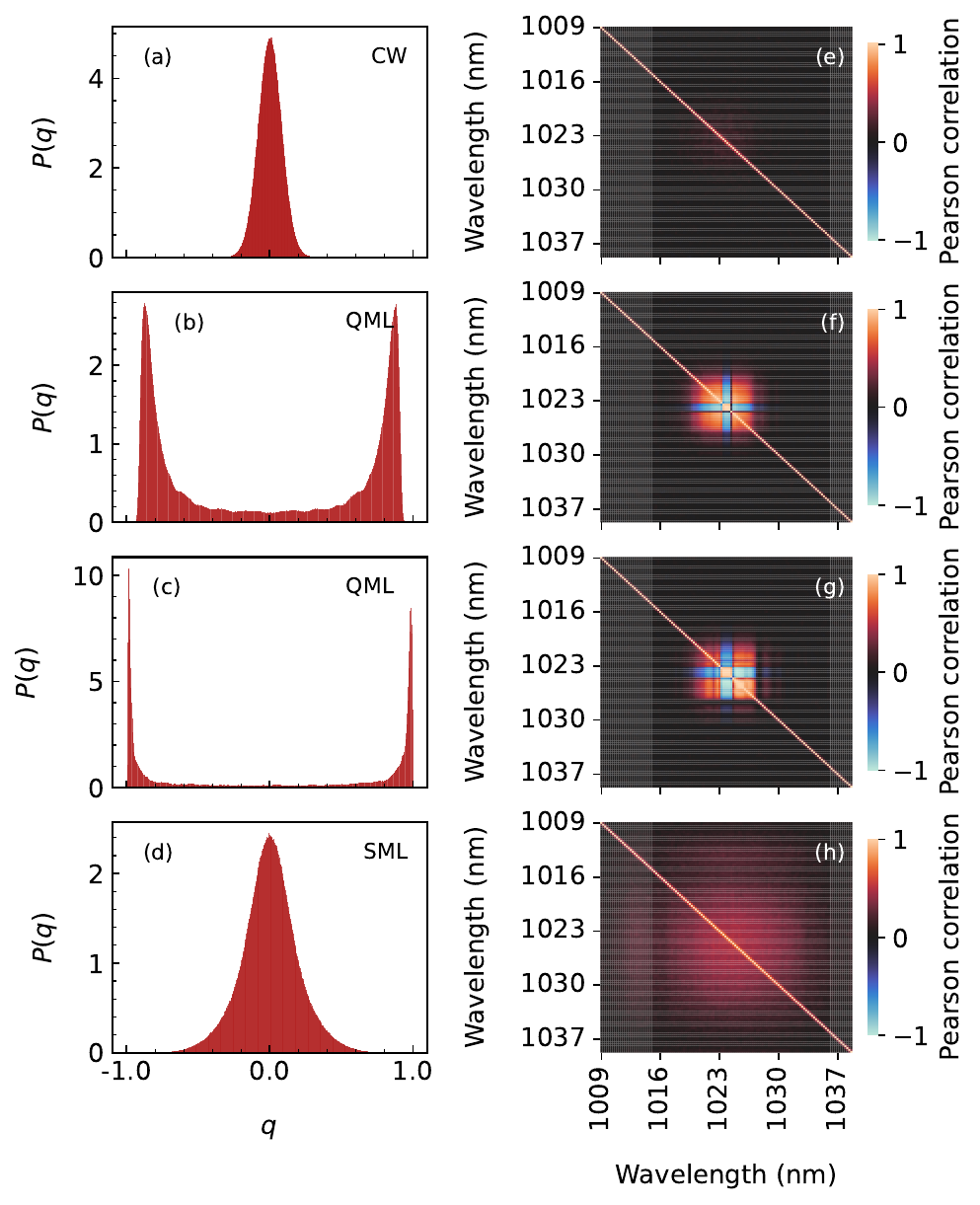}
\caption{Distribution $P(q)$ of Parisi overlap parameter $q$ (left column) and Pearson coefficient $C_{k_i k_j}$ (right) in the regimes: (a) and (e) CW (excitation current 274~mA); (b) and (f) (314~mA) and (c) and (g) (392~mA) QML with RSB; and (d) and (h) replica-symmetric SML (407~mA).}
\label{f3}
\end{figure}

Figures~\ref{f3}(a)-\ref{f3}(d) show the distribution~$P(q)$ corresponding to the photonic phases of Fig.~\ref{f2}.
In Fig.~\ref{f3}(a), the MLFL system is at the CW regime, pumped by current 274~mA (see Fig.~\ref{f2}), and the pronounced maximum at $q=0$ indicates replica symmetric behavior. 
In this case, which is the photonic analogue of the paramagnetic state with uncorrelated spins, the modes oscillate incoherently and light is emitted in the form of continuous waves.
As the excitation energy (current) raises, the optical nonlinearity starts to play an increasingly relevant role, and more modes are activated in the cavity. 
The multiple longitudinal modes compete for gain until a fully-developed SML regime sets in for large enough currents, via self- and cross-phase modulation effects~\cite{Ferman}. 
The SML emission behavior is characterized by coherent modes locked in phase, in analogy to the ferromagnetic state with spins pointing parallel, and light is emitted in the form of ultrashort pulses (typically around 275~fs).  
As shown in Fig.~\ref{f3}(d) for 407~mA, the SML regime also presents replica symmetric behavior, with $P(q)$ displaying the central maximum at $q=0$. 

However, the most interesting scenario emerges for intermediate excitation energies between the replica-symmetric CW and SML regimes. 
In this quasi-mode-locking (QML) regime, full phase coherence has not been achieved yet for all modes. 
The stochastic competition for gain introduces disorder in the MLFL system, which is reflected in the RSB double-peaked profile of $P(q)$ observed in Figs.~\ref{f3}(b) and~\ref{f3}(c), respectively for 314~mA and 392~mA.  
Indeed, as the MLFL system evolves dynamically in time under CW pumping, the partial set of modes able to achieve coherent oscillation changes non-deterministically from replica to replica, thus preventing the emitted light to take the form of the ultrashort pulses typical of the replica-symmetric SML regime. 
Instead, the RSB~QML regime of Figs.~\ref{f3}(b)-\ref{f3}(c) is not pulsed, and the overlap of intensity fluctuations in Parisi parameter signals the presence of either correlated (positive~$q$) or anticorrelated (negative~$q$) replicas.  
These findings are also compatible with the theoretical modeling~\cite{a61,a63} that explains the emergence of RSB regimes in photonic systems (see Supplemental Material~\new{\cite{sm}}).
{See, also,~\cite{sm} for an account of recent numerical studies of $P(q)$.}

At this point, it is also important to stress that we have carefully checked on the fluctuations of the pump source in order to rule out its influence on the intensity fluctuations of the MLFL system. 
Actually, by analyzing spectra emitted by the CW fiber-coupled diode laser for several current values in the range up to 800~mA, we have found that they are statistically similar, always presenting replica symmetric behavior (Supplemental Material~\new{\cite{sm}}). 
So the RSB features with double-peaked $P(q)$ in Figs.~\ref{f3}(b)-\ref{f3}(c) cannot be attributed to the pump source properties. 

The Pearson correlation $C_{k_i k_j}$ between distinct wavelengths in the same spectrum is shown in the heatmap plots of Figs.~\ref{f3}(e)-\ref{f3}(h). 
An uncorrelated scenario is noticed in Fig.~\ref{f3}(e) for 274~mA (mostly dark region with a thin red diagonal self-correlation line), consistent with the CW regime of Fig.~\ref{f3}(a). 
In contrast, Fig.~\ref{f3}(h) shows that modes are positively correlated for 407~mA (large red region) in the replica-symmetric SML picture, as expected for high phase coherence with mode locking. 
%

\begin{figure}[t]
\includegraphics[width=0.5\textwidth]{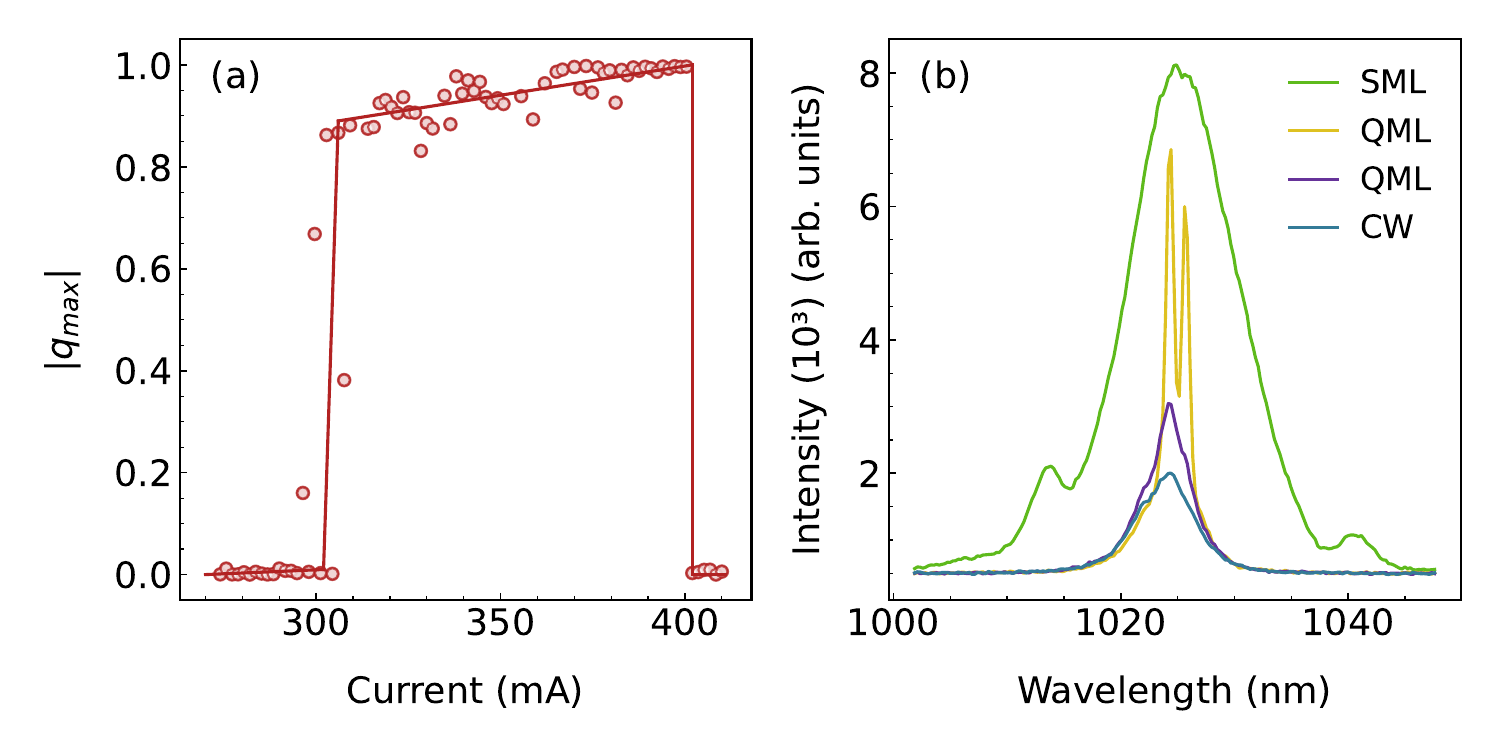}
\caption{(a) Loci $|q_{max}|$ of the maximum $P(q)$ versus current, showing the replica-symmetric-to-RSB phase transitions: CW-QML and QML-SML. (b) Typical spectra for each regime. Near the QML-SML transition (orange line, 392~mA), the peaks correspond to competing sets of modes around $\lambda_{1} = 1024.4$~nm and $\lambda_{2} = 1025.6$~nm.}
\label{f4}
\end{figure}

Interestingly, in the RSB QML regime of Figs.~\ref{f3}(f)-\ref{f3}(g), for 314~mA and 392~mA, respectively, both correlation (red) and anticorrelation (blue) of modes can be seen in the spectral range with the most prominent peaks. 
In the magnetic-to-photonic analogy, this QML picture for the mode correlations in the same replica resembles the spin-glass phase, in which up and down randomly frozen spins coexist, leading to positive and negative spin-spin correlations in the same replica (see, e.g.,~\cite{edwin,sarkar} for similar patterns of~$C_{k_i k_j}$ in photonic RSB spin-glass-like phases observed, however, in RL (not~QML) regimes). 

As the excitation energy increases from Fig.~\ref{f3}(e) to Fig.~\ref{f3}(h), the non-dark regions of the heatmaps enlarge, but with $C_{k_i k_j}$ somewhat lower than in the QML regime, evidencing the growth in the number of activated modes along with the onset of competition for gain, which ceases only in the replica-symmetric SML regime of Fig.~\ref{f3}(h), when all modes are locked in phase and oscillate coherently for large enough currents. 

\begin{figure}[t]
\includegraphics[width=0.5\textwidth]{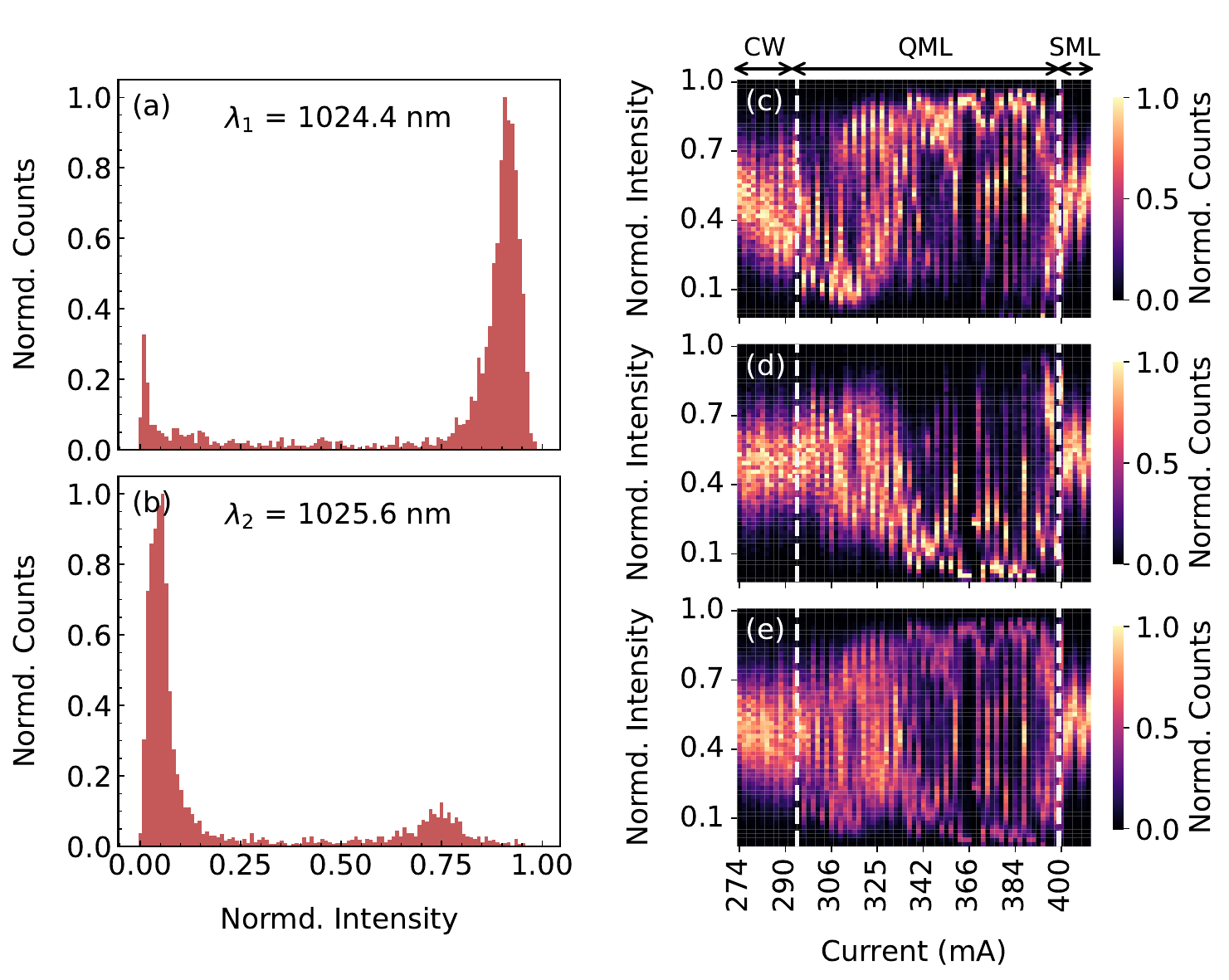}
\caption{(a)-(b) L-shaped intensity distributions corresponding to the peaks at $\lambda_{1} = 1024.4$~nm (a) and $\lambda_{2} = 1025.6$~nm (b) in Fig.~\ref{f4}(b) (orange line). 
(c)-(e) Heatmap plots of the intensity distribution in each photonic regime (CW, RSB QML, and replica-symmetric SML), associated with $\lambda_{1}$ (c), $\lambda_{2}$ (d), and superposition of $\lambda_{1}$ and $\lambda_{2}$ (e).}
\label{f5}
\end{figure}

The complex photonic behavior of the MLFL system can be further observed in Figs.~\ref{f4} and~\ref{f5}. 
First, in Fig.~\ref{f4}(a) we show the loci $q = |q_{max}|$ of points of maximum $P(q)$ as a function of the current. 
The same~sequence of regimes of Fig.~\ref{f3} can be noticed: 
a~low-current CW behavior displaying $|q_{max}| = 0$, with a sharp transition to the RSB glassy behavior in the QML regime with $|q_{max}| \approx 1$, followed by a second transition to the large-current replica-symmetric SML regime with~$|q_{max}| = 0$.  

The spectrum counterpart of this picture can be observed in Fig.~\ref{f4}(b).
We notice in both CW and SML regimes (blue and green lines, respectively) the smooth spectra, with a pronounced maximum around~1025~nm. 
The contrast with the QML spectra is evident mainly in the orange line for current 392~mA near the QML-SML transition. 
In this case, two peaks are clearly seen, corresponding to competing sets of modes with wavelengths around $\lambda_1 = 1024.4$~nm and $\lambda_2 = 1025.6$~nm. 
We also note that the CW and QML regimes display spectral ranges of $\sim 5$~nm and $\sim 6$~nm, respectively, while in the SML the bandwidth is somewhat broader, $\sim 10$-20~nm, since the longitudinal modes are locked in phase~\cite{xx}.

Figures~\ref{f5}(a) and~\ref{f5}(b) display, respectively, the intensity distribution for 392~mA, related to the peak wavelengths~$\lambda_1$ and~$\lambda_2$. 
Importantly, a close look at Fig.~\ref{f3}(g) for 392~mA shows blue strips depicting anticorrelation of modes in the range between $\lambda_1$ and~$\lambda_2$. 
In~other words, the prevalence of modes around $\lambda_1$ promotes the suppression of modes around $\lambda_2$, and vice-versa, in a gain competition scenario present in the QML regime with~RSB. 

From the statistical viewpoint, one notices in Figs.~\ref{f5}(a)-\ref{f5}(b) the L-shaped-type intensity distributions with opposite bias. 
In the photonic context, L-shaped signatures of intensity statistics can be found, e.g., in systems with many modes or multiple wave mixing displaying significant nonlinear interactions~\cite{Gonzalez:21,gao}.
It is also worth mentioning that rather distinct Gaussian-like intensity distributions are obtained in the replica-symmetric CW and SML regimes, as expected~\cite{Hh}. 

A more complete picture of the intensity flucutations in each photonic regime of the MLFL system can be observed in Figs.~\ref{f5}(c)-\ref{f5}(e) as a function of the excitation current. 
Figure~\ref{f5}(c) corresponds to the wavelength~$\lambda_1$, while Fig.~\ref{f5}(d) is related to~$\lambda_2$, and Fig.~\ref{f5}(e) shows the superposition of both Figs.~\ref{f5}(c)-\ref{f5}(d). 
The distinct regimes (CW, QML, SML) are indicated at the top of Fig.~\ref{f5}(c).
Clearly, the QML regime with RSB (central range of the plots) presents more complex patterns of intensity fluctuations. 
Close to the QML-SML transition, we note that the most likely intensity $(I)$ values occur for large-$I$ in Fig.~\ref{f5}(c) (wavelength $\lambda_1$), and for small-$I$ in Fig.~\ref{f5}(d)~($\lambda_2$). 
These results concur with Figs.~\ref{f5}(a)-\ref{f5}(b).

On the other hand, near the CW-QML transition in Figs.~\ref{f5}(c)-\ref{f5}(e) intermediate intensity values are more probable. 
We recall that in this relatively low current regime optical nonlinearities still do not play a central role in the MLFL system, and modes competition is incipient, in agreement with the CW and QML spectra shown in blue and purple lines in Fig.~\ref{f4}(b). 
%


In conclusion, here we have reported the first experimental demonstration of the replica symmetry breaking phenomenon in a 
laser system supporting standard mode-locking regime. 
Though theoretically predicted~\cite{a63}, this glassy phase remained experimentally unveiled so far. 
The relevance of photonic systems as platforms to advance the understanding of complex systems phenomena, as RSB~\cite{a81c,a59,a1c,a1,a1b}, unconventional L\'evy statistics and~extreme value events~\cite{nn3,nn4,nn6,a52,nn5}, and turbulence~\cite{a58a,a58}, can be hardly overstated. 
We hope this work helps paving~the way for exploring other striking manifestations of nontrivial RSB behavior in new classes of photonic materi\nolinebreak als. 
%

%

We thank C. B. de Ara{\'u}jo~and L.~H. Acioli for stimulating discussions. 
We also thank the financial support from the Brazilian agencies  Conselho Nacional de Desenvolvimento Cient\'{i}fico e Tecnol\'{o}gico (CNPq), Coordena\c{c}\~{a}o de Aperfei\c{c}oamento de Pessoal de N\'{i}vel Superior (CAPES), and Funda\c{c}\~{a}o de Amparo \`{a} Ci\^{e}ncia e Tecnologia do Estado de Pernambuco (FACEPE).


\bibliography{main}

\begin{thebibliography}{40}%
\makeatletter
\providecommand \@ifxundefined [1]{%
 \@ifx{#1\undefined}
}%
\providecommand \@ifnum [1]{%
 \ifnum #1\expandafter \@firstoftwo
 \else \expandafter \@secondoftwo
 \fi
}%
\providecommand \@ifx [1]{%
 \ifx #1\expandafter \@firstoftwo
 \else \expandafter \@secondoftwo
 \fi
}%
\providecommand \natexlab [1]{#1}%
\providecommand \enquote  [1]{``#1''}%
\providecommand \bibnamefont  [1]{#1}%
\providecommand \bibfnamefont [1]{#1}%
\providecommand \citenamefont [1]{#1}%
\providecommand \href@noop [0]{\@secondoftwo}%
\providecommand \href [0]{\begingroup \@sanitize@url \@href}%
\providecommand \@href[1]{\@@startlink{#1}\@@href}%
\providecommand \@@href[1]{\endgroup#1\@@endlink}%
\providecommand \@sanitize@url [0]{\catcode `\\12\catcode `\$12\catcode
  `\&12\catcode `\#12\catcode `\^12\catcode `\_12\catcode `\%12\relax}%
\providecommand \@@startlink[1]{}%
\providecommand \@@endlink[0]{}%
\providecommand \url  [0]{\begingroup\@sanitize@url \@url }%
\providecommand \@url [1]{\endgroup\@href {#1}{\urlprefix }}%
\providecommand \urlprefix  [0]{URL }%
\providecommand \Eprint [0]{\href }%
\providecommand \doibase [0]{https://doi.org/}%
\providecommand \selectlanguage [0]{\@gobble}%
\providecommand \bibinfo  [0]{\@secondoftwo}%
\providecommand \bibfield  [0]{\@secondoftwo}%
\providecommand \translation [1]{[#1]}%
\providecommand \BibitemOpen [0]{}%
\providecommand \bibitemStop [0]{}%
\providecommand \bibitemNoStop [0]{.\EOS\space}%
\providecommand \EOS [0]{\spacefactor3000\relax}%
\providecommand \BibitemShut  [1]{\csname bibitem#1\endcsname}%
\let\auto@bib@innerbib\@empty
\bibitem [{\citenamefont {Angelani}\ \emph {et~al.}(2006)\citenamefont
  {Angelani}, \citenamefont {Conti}, \citenamefont {Ruocco},\ and\
  \citenamefont {Zamponi}}]{a61}%
  \BibitemOpen
  \bibfield  {author} {\bibinfo {author} {\bibfnamefont {L.}~\bibnamefont
  {Angelani}}, \bibinfo {author} {\bibfnamefont {C.}~\bibnamefont {Conti}},
  \bibinfo {author} {\bibfnamefont {G.}~\bibnamefont {Ruocco}},\ and\ \bibinfo
  {author} {\bibfnamefont {F.}~\bibnamefont {Zamponi}},\ }\href@noop {}
  {\bibfield  {journal} {\bibinfo  {journal} {Phys. Rev. Lett.}\ }\textbf
  {\bibinfo {volume} {96}},\ \bibinfo {pages} {065702} (\bibinfo {year}
  {2006})}\BibitemShut {NoStop}%
\bibitem [{\citenamefont {Antenucci}\ \emph {et~al.}(2015)\citenamefont
  {Antenucci}, \citenamefont {Conti}, \citenamefont {Crisanti},\ and\
  \citenamefont {Leuzzi}}]{a63}%
  \BibitemOpen
  \bibfield  {author} {\bibinfo {author} {\bibfnamefont {F.}~\bibnamefont
  {Antenucci}}, \bibinfo {author} {\bibfnamefont {C.}~\bibnamefont {Conti}},
  \bibinfo {author} {\bibfnamefont {A.}~\bibnamefont {Crisanti}},\ and\
  \bibinfo {author} {\bibfnamefont {L.}~\bibnamefont {Leuzzi}},\ }\href@noop {}
  {\bibfield  {journal} {\bibinfo  {journal} {Phys. Rev. Lett.}\ }\textbf
  {\bibinfo {volume} {114}},\ \bibinfo {pages} {043901} (\bibinfo {year}
  {2015})}\BibitemShut {NoStop}%
\bibitem [{\citenamefont {Ghofraniha}\ \emph {et~al.}(2015)\citenamefont
  {Ghofraniha}, \citenamefont {Viola}, \citenamefont {Di~Maria}, \citenamefont
  {Barbarella}, \citenamefont {Gigli}, \citenamefont {Leuzzi},\ and\
  \citenamefont {Conti}}]{a66}%
  \BibitemOpen
  \bibfield  {author} {\bibinfo {author} {\bibfnamefont {N.}~\bibnamefont
  {Ghofraniha}}, \bibinfo {author} {\bibfnamefont {I.}~\bibnamefont {Viola}},
  \bibinfo {author} {\bibfnamefont {F.}~\bibnamefont {Di~Maria}}, \bibinfo
  {author} {\bibfnamefont {G.}~\bibnamefont {Barbarella}}, \bibinfo {author}
  {\bibfnamefont {G.}~\bibnamefont {Gigli}}, \bibinfo {author} {\bibfnamefont
  {L.}~\bibnamefont {Leuzzi}},\ and\ \bibinfo {author} {\bibfnamefont
  {C.}~\bibnamefont {Conti}},\ }\href@noop {} {\bibfield  {journal} {\bibinfo
  {journal} {Nat. Commun.}\ }\textbf {\bibinfo {volume} {6}},\ \bibinfo {pages}
  {6058} (\bibinfo {year} {2015})}\BibitemShut {NoStop}%
\bibitem [{\citenamefont {Gomes}\ \emph
  {et~al.}(2016{\natexlab{a}})\citenamefont {Gomes}, \citenamefont {Raposo},
  \citenamefont {Moura}, \citenamefont {Fewo}, \citenamefont {Pincheira},
  \citenamefont {Jerez}, \citenamefont {Maia},\ and\ \citenamefont
  {de~Ara\'ujo}}]{a68}%
  \BibitemOpen
  \bibfield  {author} {\bibinfo {author} {\bibfnamefont {A.~S.~L.}\
  \bibnamefont {Gomes}}, \bibinfo {author} {\bibfnamefont {E.~P.}\ \bibnamefont
  {Raposo}}, \bibinfo {author} {\bibfnamefont {A.~L.}\ \bibnamefont {Moura}},
  \bibinfo {author} {\bibfnamefont {S.~I.}\ \bibnamefont {Fewo}}, \bibinfo
  {author} {\bibfnamefont {P.~I.~R.}\ \bibnamefont {Pincheira}}, \bibinfo
  {author} {\bibfnamefont {V.}~\bibnamefont {Jerez}}, \bibinfo {author}
  {\bibfnamefont {L.~J.~Q.}\ \bibnamefont {Maia}},\ and\ \bibinfo {author}
  {\bibfnamefont {C.~B.}\ \bibnamefont {de~Ara\'ujo}},\ }\href@noop {}
  {\bibfield  {journal} {\bibinfo  {journal} {Sci. Rep.}\ }\textbf {\bibinfo
  {volume} {6}},\ \bibinfo {pages} {27987} (\bibinfo {year}
  {2016}{\natexlab{a}})}\BibitemShut {NoStop}%
\bibitem [{\citenamefont {Parisi}(2023)}]{a81c}%
  \BibitemOpen
  \bibfield  {author} {\bibinfo {author} {\bibfnamefont {G.}~\bibnamefont
  {Parisi}},\ }\href {https://doi.org/10.1103/RevModPhys.95.030501} {\bibfield
  {journal} {\bibinfo  {journal} {Rev. Mod. Phys.}\ }\textbf {\bibinfo {volume}
  {95}},\ \bibinfo {pages} {030501} (\bibinfo {year} {2023})}\BibitemShut
  {NoStop}%
\bibitem [{\citenamefont {Antenucci}(2016)}]{a59}%
  \BibitemOpen
  \bibfield  {author} {\bibinfo {author} {\bibfnamefont {F.}~\bibnamefont
  {Antenucci}},\ }\href@noop {} {\emph {\bibinfo {title} {Statistical Physics
  of Wave Interactions}}}\ (\bibinfo  {publisher} {Springer, Berlin},\ \bibinfo
  {year} {2016})\BibitemShut {NoStop}%
\bibitem [{\citenamefont {Viola}\ \emph {et~al.}(2018)\citenamefont {Viola},
  \citenamefont {Leuzzi}, \citenamefont {Conti},\ and\ \citenamefont
  {Ghofraniha}}]{a1c}%
  \BibitemOpen
  \bibfield  {author} {\bibinfo {author} {\bibfnamefont {I.}~\bibnamefont
  {Viola}}, \bibinfo {author} {\bibfnamefont {L.}~\bibnamefont {Leuzzi}},
  \bibinfo {author} {\bibfnamefont {C.}~\bibnamefont {Conti}},\ and\ \bibinfo
  {author} {\bibfnamefont {N.}~\bibnamefont {Ghofraniha}},\ }\bibinfo {title}
  {Basic physics and recent developments of organic random lasers, in
  \emph{{O}rganic {L}asers}, {M.} {A}nni and {S.} {L}attante, eds.}\ (\bibinfo
  {publisher} {Jenny Stanford Publishing, Singapore},\ \bibinfo {year} {2018})\
  p.\ \bibinfo {pages} {151}\BibitemShut {NoStop}%
\bibitem [{\citenamefont {Gomes}\ \emph {et~al.}(2023)\citenamefont {Gomes},
  \citenamefont {Moura}, \citenamefont {de~Ara\'ujo},\ and\ \citenamefont
  {Raposo}}]{a1}%
  \BibitemOpen
  \bibfield  {author} {\bibinfo {author} {\bibfnamefont {A.~S.~L.}\
  \bibnamefont {Gomes}}, \bibinfo {author} {\bibfnamefont {A.~L.}\ \bibnamefont
  {Moura}}, \bibinfo {author} {\bibfnamefont {C.~B.}\ \bibnamefont
  {de~Ara\'ujo}},\ and\ \bibinfo {author} {\bibfnamefont {E.~P.}\ \bibnamefont
  {Raposo}, \bibfnamefont {eds.}},\ }\href@noop {} {\emph {\bibinfo {title}
  {L\'evy Statistics and Spin Glass Behavior in Random Lasers –- A Photonic
  Platform for Statistical and Complex Physics}}}\ (\bibinfo  {publisher}
  {Taylor \& Francis, New York},\ \bibinfo {year} {2023})\BibitemShut {NoStop}%
\bibitem [{\citenamefont {Gomes}\ \emph {et~al.}(2021)\citenamefont {Gomes},
  \citenamefont {Moura}, \citenamefont {de~Ara{\'u}jo},\ and\ \citenamefont
  {Raposo}}]{a1b}%
  \BibitemOpen
  \bibfield  {author} {\bibinfo {author} {\bibfnamefont {A.~S.~L.}\
  \bibnamefont {Gomes}}, \bibinfo {author} {\bibfnamefont {A.~L.}\ \bibnamefont
  {Moura}}, \bibinfo {author} {\bibfnamefont {C.~B.}\ \bibnamefont
  {de~Ara{\'u}jo}},\ and\ \bibinfo {author} {\bibfnamefont {E.~P.}\
  \bibnamefont {Raposo}},\ }\href@noop {} {\bibfield  {journal} {\bibinfo
  {journal} {Prog. Quantum Electron.}\ }\textbf {\bibinfo {volume} {78}},\
  \bibinfo {pages} {100343} (\bibinfo {year} {2021})}\BibitemShut {NoStop}%
\bibitem [{\citenamefont {M\'ezard}\ \emph {et~al.}(1987)\citenamefont
  {M\'ezard}, \citenamefont {Parisi},\ and\ \citenamefont {Virasoro}}]{a80}%
  \BibitemOpen
  \bibfield  {author} {\bibinfo {author} {\bibfnamefont {M.}~\bibnamefont
  {M\'ezard}}, \bibinfo {author} {\bibfnamefont {G.}~\bibnamefont {Parisi}},\
  and\ \bibinfo {author} {\bibfnamefont {M.~A.}\ \bibnamefont {Virasoro}},\
  }\href@noop {} {\emph {\bibinfo {title} {Spin Glass Theory and Beyond}}}\
  (\bibinfo  {publisher} {World Scientific, Singapore},\ \bibinfo {year}
  {1987})\BibitemShut {NoStop}%
\bibitem [{\citenamefont {Haus}(2000)}]{ml}%
  \BibitemOpen
  \bibfield  {author} {\bibinfo {author} {\bibfnamefont {H.}~\bibnamefont
  {Haus}},\ }\bibfield  {title} {\bibinfo {title} {Mode-locking of lasers},\
  }\href {https://doi.org/10.1109/2944.902165} {\bibfield  {journal} {\bibinfo
  {journal} {IEEE J. Sel. Top. Quantum Electron.}\ }\textbf {\bibinfo {volume}
  {6}},\ \bibinfo {pages} {1173} (\bibinfo {year} {2000})}\BibitemShut
  {NoStop}%
\bibitem [{\citenamefont {Gordon}\ and\ \citenamefont {Fischer}(2003)}]{gor}%
  \BibitemOpen
  \bibfield  {author} {\bibinfo {author} {\bibfnamefont {A.}~\bibnamefont
  {Gordon}}\ and\ \bibinfo {author} {\bibfnamefont {B.}~\bibnamefont
  {Fischer}},\ }\href
  {https://doi.org/https://doi.org/10.1016/S0030-4018(03)01622-5} {\bibfield
  {journal} {\bibinfo  {journal} {Opt. Commun.}\ }\textbf {\bibinfo {volume}
  {223}},\ \bibinfo {pages} {151} (\bibinfo {year} {2003})}\BibitemShut
  {NoStop}%
\bibitem [{\citenamefont {Moura}\ \emph {et~al.}(2020)\citenamefont {Moura},
  \citenamefont {Carre{\~n}o}, \citenamefont {Pincheira}, \citenamefont {Maia},
  \citenamefont {Jerez}, \citenamefont {Raposo}, \citenamefont {Gomes},\ and\
  \citenamefont {de~Ara\'ujo}}]{bbbb}%
  \BibitemOpen
  \bibfield  {author} {\bibinfo {author} {\bibfnamefont {A.~L.}\ \bibnamefont
  {Moura}}, \bibinfo {author} {\bibfnamefont {S.~J.}\ \bibnamefont
  {Carre{\~n}o}}, \bibinfo {author} {\bibfnamefont {P.~I.~R.}\ \bibnamefont
  {Pincheira}}, \bibinfo {author} {\bibfnamefont {L.~J.~Q.}\ \bibnamefont
  {Maia}}, \bibinfo {author} {\bibfnamefont {V.}~\bibnamefont {Jerez}},
  \bibinfo {author} {\bibfnamefont {E.~P.}\ \bibnamefont {Raposo}}, \bibinfo
  {author} {\bibfnamefont {A.~S.~L.}\ \bibnamefont {Gomes}},\ and\ \bibinfo
  {author} {\bibfnamefont {C.~B.}\ \bibnamefont {de~Ara\'ujo}},\ }\href
  {https://doi.org/10.1364/AO.383477} {\bibfield  {journal} {\bibinfo
  {journal} {Appl. Opt.}\ }\textbf {\bibinfo {volume} {59}},\ \bibinfo {pages}
  {D155} (\bibinfo {year} {2020})}\BibitemShut {NoStop}%
\bibitem [{\citenamefont {Moura}\ \emph {et~al.}(2017)\citenamefont {Moura},
  \citenamefont {Pincheira}, \citenamefont {Reyna}, \citenamefont {Raposo},
  \citenamefont {Gomes},\ and\ \citenamefont {de~Ara\'ujo}}]{a71}%
  \BibitemOpen
  \bibfield  {author} {\bibinfo {author} {\bibfnamefont {A.~L.}\ \bibnamefont
  {Moura}}, \bibinfo {author} {\bibfnamefont {P.~I.~R.}\ \bibnamefont
  {Pincheira}}, \bibinfo {author} {\bibfnamefont {A.~S.}\ \bibnamefont
  {Reyna}}, \bibinfo {author} {\bibfnamefont {E.~P.}\ \bibnamefont {Raposo}},
  \bibinfo {author} {\bibfnamefont {A.~S.~L.}\ \bibnamefont {Gomes}},\ and\
  \bibinfo {author} {\bibfnamefont {C.~B.}\ \bibnamefont {de~Ara\'ujo}},\
  }\href {https://doi.org/10.1103/PhysRevLett.119.163902} {\bibfield  {journal}
  {\bibinfo  {journal} {Phys. Rev. Lett.}\ }\textbf {\bibinfo {volume} {119}},\
  \bibinfo {pages} {163902} (\bibinfo {year} {2017})}\BibitemShut {NoStop}%
\bibitem [{\citenamefont {Gomes}\ \emph
  {et~al.}(2016{\natexlab{b}})\citenamefont {Gomes}, \citenamefont {Lima},
  \citenamefont {Pincheira}, \citenamefont {Moura}, \citenamefont {Gagn\'e},
  \citenamefont {Raposo}, \citenamefont {de~Ara\'ujo},\ and\ \citenamefont
  {Kashyap}}]{anderson}%
  \BibitemOpen
  \bibfield  {author} {\bibinfo {author} {\bibfnamefont {A.~S.~L.}\
  \bibnamefont {Gomes}}, \bibinfo {author} {\bibfnamefont {B.~C.}\ \bibnamefont
  {Lima}}, \bibinfo {author} {\bibfnamefont {P.~I.~R.}\ \bibnamefont
  {Pincheira}}, \bibinfo {author} {\bibfnamefont {A.~L.}\ \bibnamefont
  {Moura}}, \bibinfo {author} {\bibfnamefont {M.}~\bibnamefont {Gagn\'e}},
  \bibinfo {author} {\bibfnamefont {E.~P.}\ \bibnamefont {Raposo}}, \bibinfo
  {author} {\bibfnamefont {C.~B.}\ \bibnamefont {de~Ara\'ujo}},\ and\ \bibinfo
  {author} {\bibfnamefont {R.}~\bibnamefont {Kashyap}},\ }\href
  {https://doi.org/10.1103/PhysRevA.94.011801} {\bibfield  {journal} {\bibinfo
  {journal} {Phys. Rev. A}\ }\textbf {\bibinfo {volume} {94}},\ \bibinfo
  {pages} {011801(R)} (\bibinfo {year} {2016}{\natexlab{b}})}\BibitemShut
  {NoStop}%
\bibitem [{\citenamefont {Coronel}\ \emph {et~al.}(2021)\citenamefont
  {Coronel}, \citenamefont {Das}, \citenamefont {Gonz\'{a}lez}, \citenamefont
  {Gomes}, \citenamefont {Margulis}, \citenamefont {von~der Weid},\ and\
  \citenamefont {Raposo}}]{edwin}%
  \BibitemOpen
  \bibfield  {author} {\bibinfo {author} {\bibfnamefont {E.}~\bibnamefont
  {Coronel}}, \bibinfo {author} {\bibfnamefont {A.}~\bibnamefont {Das}},
  \bibinfo {author} {\bibfnamefont {I.~R.~R.}\ \bibnamefont {Gonz\'{a}lez}},
  \bibinfo {author} {\bibfnamefont {A.~S.~L.}\ \bibnamefont {Gomes}}, \bibinfo
  {author} {\bibfnamefont {W.}~\bibnamefont {Margulis}}, \bibinfo {author}
  {\bibfnamefont {J.~P.}\ \bibnamefont {von~der Weid}},\ and\ \bibinfo {author}
  {\bibfnamefont {E.~P.}\ \bibnamefont {Raposo}},\ }\href
  {https://doi.org/10.1364/OE.431981} {\bibfield  {journal} {\bibinfo
  {journal} {Opt. Express}\ }\textbf {\bibinfo {volume} {29}},\ \bibinfo
  {pages} {24422} (\bibinfo {year} {2021})}\BibitemShut {NoStop}%
\bibitem [{\citenamefont {M{\'e}lo}\ \emph {et~al.}(2018)\citenamefont
  {M{\'e}lo}, \citenamefont {Palacios}, \citenamefont {Carelli}, \citenamefont
  {Acioli}, \citenamefont {Leite},\ and\ \citenamefont {de~Miranda}}]{Lucas18}%
  \BibitemOpen
  \bibfield  {author} {\bibinfo {author} {\bibfnamefont {L.~B.~A.}\
  \bibnamefont {M{\'e}lo}}, \bibinfo {author} {\bibfnamefont {G.~F.~R.}\
  \bibnamefont {Palacios}}, \bibinfo {author} {\bibfnamefont {P.~V.}\
  \bibnamefont {Carelli}}, \bibinfo {author} {\bibfnamefont {L.~H.}\
  \bibnamefont {Acioli}}, \bibinfo {author} {\bibfnamefont {J.~R.~R.}\
  \bibnamefont {Leite}},\ and\ \bibinfo {author} {\bibfnamefont {M.~H.~G.}\
  \bibnamefont {de~Miranda}},\ }\href@noop {} {\bibfield  {journal} {\bibinfo
  {journal} {Opt. Express}\ }\textbf {\bibinfo {volume} {26}},\ \bibinfo
  {pages} {13686} (\bibinfo {year} {2018})}\BibitemShut {NoStop}%
\bibitem [{\citenamefont {Li}\ \emph {et~al.}(2010)\citenamefont {Li},
  \citenamefont {Wai},\ and\ \citenamefont {Kutz}}]{li2010geometrical}%
  \BibitemOpen
  \bibfield  {author} {\bibinfo {author} {\bibfnamefont {F.}~\bibnamefont
  {Li}}, \bibinfo {author} {\bibfnamefont {P.~K.~A.}\ \bibnamefont {Wai}},\
  and\ \bibinfo {author} {\bibfnamefont {J.~N.}\ \bibnamefont {Kutz}},\
  }\href@noop {} {\bibfield  {journal} {\bibinfo  {journal} {J. Opt. Soc. Am.
  B}\ }\textbf {\bibinfo {volume} {27}},\ \bibinfo {pages} {2068} (\bibinfo
  {year} {2010})}\BibitemShut {NoStop}%
\bibitem [{\citenamefont {Campos}\ \emph {et~al.}(2020)\citenamefont {Campos},
  \citenamefont {M{\'e}lo}, \citenamefont {de~S.~Cavalcante}, \citenamefont
  {Acioli},\ and\ \citenamefont {{de Miranda}}}]{Cecilia19}%
  \BibitemOpen
  \bibfield  {author} {\bibinfo {author} {\bibfnamefont {C.~L. A.~V.}\
  \bibnamefont {Campos}}, \bibinfo {author} {\bibfnamefont {L.~B.}\
  \bibnamefont {M{\'e}lo}}, \bibinfo {author} {\bibfnamefont {H.~L.}\
  \bibnamefont {de~S.~Cavalcante}}, \bibinfo {author} {\bibfnamefont {L.~H.}\
  \bibnamefont {Acioli}},\ and\ \bibinfo {author} {\bibfnamefont {M.~H.}\
  \bibnamefont {{de Miranda}}},\ }\href@noop {} {\bibfield  {journal} {\bibinfo
   {journal} {Opt. Commun.}\ }\textbf {\bibinfo {volume} {461}},\ \bibinfo
  {pages} {125154} (\bibinfo {year} {2020})}\BibitemShut {NoStop}%
\bibitem [{\citenamefont {Hofer}\ \emph {et~al.}(1991)\citenamefont {Hofer},
  \citenamefont {Fermann}, \citenamefont {Haberl}, \citenamefont {Ober},\ and\
  \citenamefont {Schmidt}}]{Ferman}%
  \BibitemOpen
  \bibfield  {author} {\bibinfo {author} {\bibfnamefont {M.}~\bibnamefont
  {Hofer}}, \bibinfo {author} {\bibfnamefont {M.~E.}\ \bibnamefont {Fermann}},
  \bibinfo {author} {\bibfnamefont {F.}~\bibnamefont {Haberl}}, \bibinfo
  {author} {\bibfnamefont {M.~H.}\ \bibnamefont {Ober}},\ and\ \bibinfo
  {author} {\bibfnamefont {A.~J.}\ \bibnamefont {Schmidt}},\ }\href@noop {}
  {\bibfield  {journal} {\bibinfo  {journal} {Opt. Lett.}\ }\textbf {\bibinfo
  {volume} {16}},\ \bibinfo {pages} {502} (\bibinfo {year} {1991})}\BibitemShut
  {NoStop}%
\bibitem [{sm()}]{sm}%
  \BibitemOpen
  \href@noop {} {\bibinfo  {journal} {\new{See Supplemental Material [url] for
  further experimental details and characterization of the RSB QML phase, which
  includes Refs.~\cite{Campos21,rev1,rev2b,rev2,rev3,rev4,ti_safira}}}\
  }\BibitemShut {NoStop}%
\bibitem [{\citenamefont {Sarkar}\ \emph {et~al.}(2020)\citenamefont {Sarkar},
  \citenamefont {Bhaktha},\ and\ \citenamefont {Andreasen}}]{sarkar}%
  \BibitemOpen
\bibfield  {journal} {  }\bibfield  {author} {\bibinfo {author} {\bibfnamefont
  {A.}~\bibnamefont {Sarkar}}, \bibinfo {author} {\bibfnamefont {B.~N.~S.}\
  \bibnamefont {Bhaktha}},\ and\ \bibinfo {author} {\bibfnamefont
  {J.}~\bibnamefont {Andreasen}},\ }\href@noop {} {\bibfield  {journal}
  {\bibinfo  {journal} {Sci. Rep.}\ }\textbf {\bibinfo {volume} {10}},\
  \bibinfo {pages} {2628} (\bibinfo {year} {2020})}\BibitemShut {NoStop}%
\bibitem [{\citenamefont {Ippen}\ \emph {et~al.}(2003)\citenamefont {Ippen},
  \citenamefont {Shank},\ and\ \citenamefont {Dienes}}]{xx}%
  \BibitemOpen
  \bibfield  {author} {\bibinfo {author} {\bibfnamefont {E.}~\bibnamefont
  {Ippen}}, \bibinfo {author} {\bibfnamefont {C.}~\bibnamefont {Shank}},\ and\
  \bibinfo {author} {\bibfnamefont {A.}~\bibnamefont {Dienes}},\ }\href
  {https://doi.org/10.1063/1.1654406} {\bibfield  {journal} {\bibinfo
  {journal} {Appl. Phys. Lett.}\ }\textbf {\bibinfo {volume} {21}},\ \bibinfo
  {pages} {348} (\bibinfo {year} {2003})}\BibitemShut {NoStop}%
\bibitem [{\citenamefont {Gonz\'{a}lez}\ \emph {et~al.}(2021)\citenamefont
  {Gonz\'{a}lez}, \citenamefont {Pincheira}, \citenamefont {Mac\^{e}do},
  \citenamefont {de~S.~Menezes}, \citenamefont {Gomes},\ and\ \citenamefont
  {Raposo}}]{Gonzalez:21}%
  \BibitemOpen
  \bibfield  {author} {\bibinfo {author} {\bibfnamefont {I.~R.~R.}\
  \bibnamefont {Gonz\'{a}lez}}, \bibinfo {author} {\bibfnamefont {P.~I.~R.}\
  \bibnamefont {Pincheira}}, \bibinfo {author} {\bibfnamefont {A.~M.~S.}\
  \bibnamefont {Mac\^{e}do}}, \bibinfo {author} {\bibfnamefont
  {L.}~\bibnamefont {de~S.~Menezes}}, \bibinfo {author} {\bibfnamefont
  {A.~S.~L.}\ \bibnamefont {Gomes}},\ and\ \bibinfo {author} {\bibfnamefont
  {E.~P.}\ \bibnamefont {Raposo}},\ }\href
  {https://doi.org/10.1364/JOSAB.433317} {\bibfield  {journal} {\bibinfo
  {journal} {J. Opt. Soc. Am. B}\ }\textbf {\bibinfo {volume} {38}},\ \bibinfo
  {pages} {2391} (\bibinfo {year} {2021})}\BibitemShut {NoStop}%
\bibitem [{\citenamefont {Gao}\ \emph {et~al.}(2020)\citenamefont {Gao},
  \citenamefont {Wu}, \citenamefont {Cao}, \citenamefont {Wabnitz},\ and\
  \citenamefont {Zhu}}]{gao}%
  \BibitemOpen
  \bibfield  {author} {\bibinfo {author} {\bibfnamefont {L.}~\bibnamefont
  {Gao}}, \bibinfo {author} {\bibfnamefont {Q.}~\bibnamefont {Wu}}, \bibinfo
  {author} {\bibfnamefont {Y.}~\bibnamefont {Cao}}, \bibinfo {author}
  {\bibfnamefont {S.}~\bibnamefont {Wabnitz}},\ and\ \bibinfo {author}
  {\bibfnamefont {T.}~\bibnamefont {Zhu}},\ }\href@noop {} {\bibfield
  {journal} {\bibinfo  {journal} {J. Phys. Photonics}\ }\textbf {\bibinfo
  {volume} {2}},\ \bibinfo {pages} {032004} (\bibinfo {year}
  {2020})}\BibitemShut {NoStop}%
\bibitem [{\citenamefont {Haken}(1985)}]{Hh}%
  \BibitemOpen
  \bibfield  {author} {\bibinfo {author} {\bibfnamefont {H.}~\bibnamefont
  {Haken}},\ }\href@noop {} {\emph {\bibinfo {title} {Light}}}\ (\bibinfo
  {publisher} {North-Holland, Amsterdam},\ \bibinfo {year} {1985})\BibitemShut
  {NoStop}%
\bibitem [{\citenamefont {Barthelemy}\ \emph {et~al.}(2008)\citenamefont
  {Barthelemy}, \citenamefont {Bertolotti},\ and\ \citenamefont
  {Wiersma}}]{nn3}%
  \BibitemOpen
  \bibfield  {author} {\bibinfo {author} {\bibfnamefont {P.}~\bibnamefont
  {Barthelemy}}, \bibinfo {author} {\bibfnamefont {J.}~\bibnamefont
  {Bertolotti}},\ and\ \bibinfo {author} {\bibfnamefont {D.}~\bibnamefont
  {Wiersma}},\ }\bibfield  {title} {\bibinfo {title} {A {L}\'evy flight for
  light},\ }\href {https://doi.org/https://doi.org/10.1038/nature06948}
  {\bibfield  {journal} {\bibinfo  {journal} {Nature (London)}\ }\textbf
  {\bibinfo {volume} {453}},\ \bibinfo {pages} {495} (\bibinfo {year}
  {2008})}\BibitemShut {NoStop}%
\bibitem [{\citenamefont {Raposo}\ and\ \citenamefont {Gomes}(2015)}]{nn4}%
  \BibitemOpen
  \bibfield  {author} {\bibinfo {author} {\bibfnamefont {E.~P.}\ \bibnamefont
  {Raposo}}\ and\ \bibinfo {author} {\bibfnamefont {A.~S.~L.}\ \bibnamefont
  {Gomes}},\ }\href {https://doi.org/10.1103/PhysRevA.91.043827} {\bibfield
  {journal} {\bibinfo  {journal} {Phys. Rev. A}\ }\textbf {\bibinfo {volume}
  {91}},\ \bibinfo {pages} {043827} (\bibinfo {year} {2015})}\BibitemShut
  {NoStop}%
\bibitem [{\citenamefont {de~Ara{\'u}jo}\ \emph {et~al.}(2017)\citenamefont
  {de~Ara{\'u}jo}, \citenamefont {Gomes},\ and\ \citenamefont {Raposo}}]{nn6}%
  \BibitemOpen
  \bibfield  {author} {\bibinfo {author} {\bibfnamefont {C.~B.}\ \bibnamefont
  {de~Ara{\'u}jo}}, \bibinfo {author} {\bibfnamefont {A.~S.~L.}\ \bibnamefont
  {Gomes}},\ and\ \bibinfo {author} {\bibfnamefont {E.~P.}\ \bibnamefont
  {Raposo}},\ }\href {https://doi.org/10.3390/app7070644} {\bibfield  {journal}
  {\bibinfo  {journal} {Appl. Sci.}\ }\textbf {\bibinfo {volume} {7}},\
  \bibinfo {pages} {644} (\bibinfo {year} {2017})}\BibitemShut {NoStop}%
\bibitem [{\citenamefont {Uppu}\ and\ \citenamefont {Mujumdar}(2015)}]{a52}%
  \BibitemOpen
  \bibfield  {author} {\bibinfo {author} {\bibfnamefont {R.}~\bibnamefont
  {Uppu}}\ and\ \bibinfo {author} {\bibfnamefont {S.}~\bibnamefont
  {Mujumdar}},\ }\href {https://doi.org/10.1364/OL.40.005046} {\bibfield
  {journal} {\bibinfo  {journal} {Opt. Lett.}\ }\textbf {\bibinfo {volume}
  {40}},\ \bibinfo {pages} {5046} (\bibinfo {year} {2015})}\BibitemShut
  {NoStop}%
\bibitem [{\citenamefont {Lima}\ \emph {et~al.}(2017)\citenamefont {Lima},
  \citenamefont {Pincheira}, \citenamefont {Raposo}, \citenamefont
  {de~S.~Menezes}, \citenamefont {de~Ara\'ujo}, \citenamefont {Gomes},\ and\
  \citenamefont {Kashyap}}]{nn5}%
  \BibitemOpen
  \bibfield  {author} {\bibinfo {author} {\bibfnamefont {B.~C.}\ \bibnamefont
  {Lima}}, \bibinfo {author} {\bibfnamefont {P.~I.~R.}\ \bibnamefont
  {Pincheira}}, \bibinfo {author} {\bibfnamefont {E.~P.}\ \bibnamefont
  {Raposo}}, \bibinfo {author} {\bibfnamefont {L.}~\bibnamefont
  {de~S.~Menezes}}, \bibinfo {author} {\bibfnamefont {C.~B.}\ \bibnamefont
  {de~Ara\'ujo}}, \bibinfo {author} {\bibfnamefont {A.~S.~L.}\ \bibnamefont
  {Gomes}},\ and\ \bibinfo {author} {\bibfnamefont {R.}~\bibnamefont
  {Kashyap}},\ }\href {https://doi.org/10.1103/PhysRevA.96.013834} {\bibfield
  {journal} {\bibinfo  {journal} {Phys. Rev. A}\ }\textbf {\bibinfo {volume}
  {96}},\ \bibinfo {pages} {013834} (\bibinfo {year} {2017})}\BibitemShut
  {NoStop}%
\bibitem [{\citenamefont {Turitsyna}\ \emph {et~al.}(2013)\citenamefont
  {Turitsyna}, \citenamefont {Smirnov}, \citenamefont {Sugavanam},
  \citenamefont {Tarasov}, \citenamefont {Shu}, \citenamefont {Babin},
  \citenamefont {Podivilov}, \citenamefont {Churkin}, \citenamefont
  {Falkovich},\ and\ \citenamefont {Turitsyn}}]{a58a}%
  \BibitemOpen
  \bibfield  {author} {\bibinfo {author} {\bibfnamefont {E.~G.}\ \bibnamefont
  {Turitsyna}}, \bibinfo {author} {\bibfnamefont {S.~V.}\ \bibnamefont
  {Smirnov}}, \bibinfo {author} {\bibfnamefont {S.}~\bibnamefont {Sugavanam}},
  \bibinfo {author} {\bibfnamefont {N.}~\bibnamefont {Tarasov}}, \bibinfo
  {author} {\bibfnamefont {X.}~\bibnamefont {Shu}}, \bibinfo {author}
  {\bibfnamefont {S.~A.}\ \bibnamefont {Babin}}, \bibinfo {author}
  {\bibfnamefont {E.~V.}\ \bibnamefont {Podivilov}}, \bibinfo {author}
  {\bibfnamefont {D.~V.}\ \bibnamefont {Churkin}}, \bibinfo {author}
  {\bibfnamefont {G.}~\bibnamefont {Falkovich}},\ and\ \bibinfo {author}
  {\bibfnamefont {S.~K.}\ \bibnamefont {Turitsyn}},\ }\href@noop {} {\bibfield
  {journal} {\bibinfo  {journal} {Nat. Photonics}\ }\textbf {\bibinfo {volume}
  {7}},\ \bibinfo {pages} {783} (\bibinfo {year} {2013})}\BibitemShut {NoStop}%
\bibitem [{\citenamefont {Gonz\'alez}\ \emph {et~al.}(2017)\citenamefont
  {Gonz\'alez}, \citenamefont {Lima}, \citenamefont {Pincheira}, \citenamefont
  {Brum}, \citenamefont {Mac\^edo}, \citenamefont {Vasconcelos}, \citenamefont
  {de~S.~Menezes}, \citenamefont {Raposo}, \citenamefont {Gomes},\ and\
  \citenamefont {Kashyap}}]{a58}%
  \BibitemOpen
  \bibfield  {author} {\bibinfo {author} {\bibfnamefont {I.~R.~R.}\
  \bibnamefont {Gonz\'alez}}, \bibinfo {author} {\bibfnamefont {B.~C.}\
  \bibnamefont {Lima}}, \bibinfo {author} {\bibfnamefont {P.~I.~R.}\
  \bibnamefont {Pincheira}}, \bibinfo {author} {\bibfnamefont {A.~A.}\
  \bibnamefont {Brum}}, \bibinfo {author} {\bibfnamefont {A.~M.~S.}\
  \bibnamefont {Mac\^edo}}, \bibinfo {author} {\bibfnamefont {G.~L.}\
  \bibnamefont {Vasconcelos}}, \bibinfo {author} {\bibfnamefont
  {L.}~\bibnamefont {de~S.~Menezes}}, \bibinfo {author} {\bibfnamefont {E.~P.}\
  \bibnamefont {Raposo}}, \bibinfo {author} {\bibfnamefont {A.~S.~L.}\
  \bibnamefont {Gomes}},\ and\ \bibinfo {author} {\bibfnamefont
  {R.}~\bibnamefont {Kashyap}},\ }\href@noop {} {\bibfield  {journal} {\bibinfo
   {journal} {Nat. Commun.}\ }\textbf {\bibinfo {volume} {8}},\ \bibinfo
  {pages} {15731} (\bibinfo {year} {2017})}\BibitemShut {NoStop}%
\bibitem [{\citenamefont {Campos}\ \emph {et~al.}(2021)\citenamefont {Campos},
  \citenamefont {Acioli},\ and\ \citenamefont {de~Miranda}}]{Campos21}%
  \BibitemOpen
  \bibfield  {author} {\bibinfo {author} {\bibfnamefont {C.~L. A.~V.}\
  \bibnamefont {Campos}}, \bibinfo {author} {\bibfnamefont {L.~H.}\
  \bibnamefont {Acioli}},\ and\ \bibinfo {author} {\bibfnamefont {M.~H.~G.}\
  \bibnamefont {de~Miranda}},\ }\href@noop {} {\bibfield  {journal} {\bibinfo
  {journal} {Laser Phys.}\ }\textbf {\bibinfo {volume} {31}},\ \bibinfo {pages}
  {085103} (\bibinfo {year} {2021})}\BibitemShut {NoStop}%
\bibitem [{\citenamefont {Antenucci}\ \emph {et~al.}(2016)\citenamefont
  {Antenucci}, \citenamefont {Crisanti}, \citenamefont {Ibáñez-Berganza},
  \citenamefont {Marruzzo},\ and\ \citenamefont {Leuzzi}}]{rev1}%
  \BibitemOpen
  \bibfield  {author} {\bibinfo {author} {\bibfnamefont {F.}~\bibnamefont
  {Antenucci}}, \bibinfo {author} {\bibfnamefont {A.}~\bibnamefont {Crisanti}},
  \bibinfo {author} {\bibfnamefont {M.}~\bibnamefont {Ibáñez-Berganza}},
  \bibinfo {author} {\bibfnamefont {A.}~\bibnamefont {Marruzzo}},\ and\
  \bibinfo {author} {\bibfnamefont {L.}~\bibnamefont {Leuzzi}},\ }\href@noop {}
  {\bibfield  {journal} {\bibinfo  {journal} {Phil. Mag.}\ }\textbf {\bibinfo
  {volume} {96}},\ \bibinfo {pages} {704} (\bibinfo {year} {2016})}\BibitemShut
  {NoStop}%
\bibitem [{\citenamefont {Gradenigo}\ \emph {et~al.}(2020)\citenamefont
  {Gradenigo}, \citenamefont {Antenucci},\ and\ \citenamefont
  {Leuzzi}}]{rev2b}%
  \BibitemOpen
  \bibfield  {author} {\bibinfo {author} {\bibfnamefont {G.}~\bibnamefont
  {Gradenigo}}, \bibinfo {author} {\bibfnamefont {F.}~\bibnamefont
  {Antenucci}},\ and\ \bibinfo {author} {\bibfnamefont {L.}~\bibnamefont
  {Leuzzi}},\ }\href {https://doi.org/10.1103/PhysRevResearch.2.023399}
  {\bibfield  {journal} {\bibinfo  {journal} {Phys. Rev. Res.}\ }\textbf
  {\bibinfo {volume} {2}},\ \bibinfo {pages} {023399} (\bibinfo {year}
  {2020})}\BibitemShut {NoStop}%
\bibitem [{\citenamefont {Niedda}\ \emph {et~al.}(2023)\citenamefont {Niedda},
  \citenamefont {Gradenigo}, \citenamefont {Leuzzi},\ and\ \citenamefont
  {Parisi}}]{rev2}%
  \BibitemOpen
  \bibfield  {author} {\bibinfo {author} {\bibfnamefont {J.}~\bibnamefont
  {Niedda}}, \bibinfo {author} {\bibfnamefont {G.}~\bibnamefont {Gradenigo}},
  \bibinfo {author} {\bibfnamefont {L.}~\bibnamefont {Leuzzi}},\ and\ \bibinfo
  {author} {\bibfnamefont {G.}~\bibnamefont {Parisi}},\ }\href@noop {}
  {\bibfield  {journal} {\bibinfo  {journal} {SciPost Phys.}\ }\textbf
  {\bibinfo {volume} {14}},\ \bibinfo {pages} {144} (\bibinfo {year}
  {2023})}\BibitemShut {NoStop}%
\bibitem [{\citenamefont {Palacios}\ \emph {et~al.}(2023)\citenamefont
  {Palacios}, \citenamefont {Siqueira}, \citenamefont {de~Ara\'ujo},
  \citenamefont {Gomes},\ and\ \citenamefont {Raposo}}]{rev3}%
  \BibitemOpen
  \bibfield  {author} {\bibinfo {author} {\bibfnamefont {G.}~\bibnamefont
  {Palacios}}, \bibinfo {author} {\bibfnamefont {A.~C.~A.}\ \bibnamefont
  {Siqueira}}, \bibinfo {author} {\bibfnamefont {C.~B.}\ \bibnamefont
  {de~Ara\'ujo}}, \bibinfo {author} {\bibfnamefont {A.~S.~L.}\ \bibnamefont
  {Gomes}},\ and\ \bibinfo {author} {\bibfnamefont {E.~P.}\ \bibnamefont
  {Raposo}},\ }\href@noop {} {\bibfield  {journal} {\bibinfo  {journal} {Phys.
  Rev. A}\ }\textbf {\bibinfo {volume} {107}},\ \bibinfo {pages} {063510}
  (\bibinfo {year} {2023})}\BibitemShut {NoStop}%
\bibitem [{\citenamefont {Hackenbroich}\ \emph {et~al.}(2002)\citenamefont
  {Hackenbroich}, \citenamefont {Viviescas},\ and\ \citenamefont
  {Haake}}]{rev4}%
  \BibitemOpen
  \bibfield  {author} {\bibinfo {author} {\bibfnamefont {G.}~\bibnamefont
  {Hackenbroich}}, \bibinfo {author} {\bibfnamefont {C.}~\bibnamefont
  {Viviescas}},\ and\ \bibinfo {author} {\bibfnamefont {F.}~\bibnamefont
  {Haake}},\ }\href@noop {} {\bibfield  {journal} {\bibinfo  {journal} {Phys.
  Rev. Lett.}\ }\textbf {\bibinfo {volume} {89}},\ \bibinfo {pages} {083902}
  (\bibinfo {year} {2002})}\BibitemShut {NoStop}%
\bibitem [{\citenamefont {de~Ara\'ujo}\ \emph {et~al.}(2014)\citenamefont
  {de~Ara\'ujo}, \citenamefont {Roslund}, \citenamefont {Cai}, \citenamefont
  {Ferrini}, \citenamefont {Fabre},\ and\ \citenamefont {Treps}}]{ti_safira}%
  \BibitemOpen
  \bibfield  {author} {\bibinfo {author} {\bibfnamefont {R.~M.}\ \bibnamefont
  {de~Ara\'ujo}}, \bibinfo {author} {\bibfnamefont {J.}~\bibnamefont
  {Roslund}}, \bibinfo {author} {\bibfnamefont {Y.}~\bibnamefont {Cai}},
  \bibinfo {author} {\bibfnamefont {G.}~\bibnamefont {Ferrini}}, \bibinfo
  {author} {\bibfnamefont {C.}~\bibnamefont {Fabre}},\ and\ \bibinfo {author}
  {\bibfnamefont {N.}~\bibnamefont {Treps}},\ }\href@noop {} {\bibfield
  {journal} {\bibinfo  {journal} {Phys. Rev. A}\ }\textbf {\bibinfo {volume}
  {89}},\ \bibinfo {pages} {053828} (\bibinfo {year} {2014})}\BibitemShut
  {NoStop}%
\end{thebibliography}%


\begin{thebibliography}{12}%
\makeatletter
\providecommand \@ifxundefined [1]{%
 \@ifx{#1\undefined}
}%
\providecommand \@ifnum [1]{%
 \ifnum #1\expandafter \@firstoftwo
 \else \expandafter \@secondoftwo
 \fi
}%
\providecommand \@ifx [1]{%
 \ifx #1\expandafter \@firstoftwo
 \else \expandafter \@secondoftwo
 \fi
}%
\providecommand \natexlab [1]{#1}%
\providecommand \enquote  [1]{``#1''}%
\providecommand \bibnamefont  [1]{#1}%
\providecommand \bibfnamefont [1]{#1}%
\providecommand \citenamefont [1]{#1}%
\providecommand \href@noop [0]{\@secondoftwo}%
\providecommand \href [0]{\begingroup \@sanitize@url \@href}%
\providecommand \@href[1]{\@@startlink{#1}\@@href}%
\providecommand \@@href[1]{\endgroup#1\@@endlink}%
\providecommand \@sanitize@url [0]{\catcode `\\12\catcode `\$12\catcode
  `\&12\catcode `\#12\catcode `\^12\catcode `\_12\catcode `\%12\relax}%
\providecommand \@@startlink[1]{}%
\providecommand \@@endlink[0]{}%
\providecommand \url  [0]{\begingroup\@sanitize@url \@url }%
\providecommand \@url [1]{\endgroup\@href {#1}{\urlprefix }}%
\providecommand \urlprefix  [0]{URL }%
\providecommand \Eprint [0]{\href }%
\providecommand \doibase [0]{https://doi.org/}%
\providecommand \selectlanguage [0]{\@gobble}%
\providecommand \bibinfo  [0]{\@secondoftwo}%
\providecommand \bibfield  [0]{\@secondoftwo}%
\providecommand \translation [1]{[#1]}%
\providecommand \BibitemOpen [0]{}%
\providecommand \bibitemStop [0]{}%
\providecommand \bibitemNoStop [0]{.\EOS\space}%
\providecommand \EOS [0]{\spacefactor3000\relax}%
\providecommand \BibitemShut  [1]{\csname bibitem#1\endcsname}%
\let\auto@bib@innerbib\@empty
\bibitem [{\citenamefont {M{\'e}lo}\ \emph {et~al.}(2018)\citenamefont
  {M{\'e}lo}, \citenamefont {Palacios}, \citenamefont {Carelli}, \citenamefont
  {Acioli}, \citenamefont {Leite},\ and\ \citenamefont {de~Miranda}}]{Lucas18}%
  \BibitemOpen
  \bibfield  {author} {\bibinfo {author} {\bibfnamefont {L.~B.~A.}\
  \bibnamefont {M{\'e}lo}}, \bibinfo {author} {\bibfnamefont {G.~F.~R.}\
  \bibnamefont {Palacios}}, \bibinfo {author} {\bibfnamefont {P.~V.}\
  \bibnamefont {Carelli}}, \bibinfo {author} {\bibfnamefont {L.~H.}\
  \bibnamefont {Acioli}}, \bibinfo {author} {\bibfnamefont {J.~R.~R.}\
  \bibnamefont {Leite}},\ and\ \bibinfo {author} {\bibfnamefont {M.~H.~G.}\
  \bibnamefont {de~Miranda}},\ }\href@noop {} {\bibfield  {journal} {\bibinfo
  {journal} {Opt. Express}\ }\textbf {\bibinfo {volume} {26}},\ \bibinfo
  {pages} {13686} (\bibinfo {year} {2018})}\BibitemShut {NoStop}%
\bibitem [{\citenamefont {Campos}\ \emph {et~al.}(2021)\citenamefont {Campos},
  \citenamefont {Acioli},\ and\ \citenamefont {de~Miranda}}]{Campos21}%
  \BibitemOpen
  \bibfield  {author} {\bibinfo {author} {\bibfnamefont {C.~L. A.~V.}\
  \bibnamefont {Campos}}, \bibinfo {author} {\bibfnamefont {L.~H.}\
  \bibnamefont {Acioli}},\ and\ \bibinfo {author} {\bibfnamefont {M.~H.~G.}\
  \bibnamefont {de~Miranda}},\ }\href@noop {} {\bibfield  {journal} {\bibinfo
  {journal} {Laser Phys.}\ }\textbf {\bibinfo {volume} {31}},\ \bibinfo {pages}
  {085103} (\bibinfo {year} {2021})}\BibitemShut {NoStop}%
\bibitem [{\citenamefont {Campos}\ \emph {et~al.}(2020)\citenamefont {Campos},
  \citenamefont {M{\'e}lo}, \citenamefont {de~S.~Cavalcante}, \citenamefont
  {Acioli},\ and\ \citenamefont {{de Miranda}}}]{Cecilia19}%
  \BibitemOpen
  \bibfield  {author} {\bibinfo {author} {\bibfnamefont {C.~L. A.~V.}\
  \bibnamefont {Campos}}, \bibinfo {author} {\bibfnamefont {L.~B.}\
  \bibnamefont {M{\'e}lo}}, \bibinfo {author} {\bibfnamefont {H.~L.}\
  \bibnamefont {de~S.~Cavalcante}}, \bibinfo {author} {\bibfnamefont {L.~H.}\
  \bibnamefont {Acioli}},\ and\ \bibinfo {author} {\bibfnamefont {M.~H.}\
  \bibnamefont {{de Miranda}}},\ }\href@noop {} {\bibfield  {journal} {\bibinfo
   {journal} {Opt. Commun.}\ }\textbf {\bibinfo {volume} {461}},\ \bibinfo
  {pages} {125154} (\bibinfo {year} {2020})}\BibitemShut {NoStop}%
\bibitem [{\citenamefont {Hofer}\ \emph {et~al.}(1991)\citenamefont {Hofer},
  \citenamefont {Fermann}, \citenamefont {Haberl}, \citenamefont {Ober},\ and\
  \citenamefont {Schmidt}}]{Ferman}%
  \BibitemOpen
  \bibfield  {author} {\bibinfo {author} {\bibfnamefont {M.}~\bibnamefont
  {Hofer}}, \bibinfo {author} {\bibfnamefont {M.~E.}\ \bibnamefont {Fermann}},
  \bibinfo {author} {\bibfnamefont {F.}~\bibnamefont {Haberl}}, \bibinfo
  {author} {\bibfnamefont {M.~H.}\ \bibnamefont {Ober}},\ and\ \bibinfo
  {author} {\bibfnamefont {A.~J.}\ \bibnamefont {Schmidt}},\ }\href@noop {}
  {\bibfield  {journal} {\bibinfo  {journal} {Opt. Lett.}\ }\textbf {\bibinfo
  {volume} {16}},\ \bibinfo {pages} {502} (\bibinfo {year} {1991})}\BibitemShut
  {NoStop}%
\bibitem [{\citenamefont {Angelani}\ \emph {et~al.}(2006)\citenamefont
  {Angelani}, \citenamefont {Conti}, \citenamefont {Ruocco},\ and\
  \citenamefont {Zamponi}}]{a61}%
  \BibitemOpen
  \bibfield  {author} {\bibinfo {author} {\bibfnamefont {L.}~\bibnamefont
  {Angelani}}, \bibinfo {author} {\bibfnamefont {C.}~\bibnamefont {Conti}},
  \bibinfo {author} {\bibfnamefont {G.}~\bibnamefont {Ruocco}},\ and\ \bibinfo
  {author} {\bibfnamefont {F.}~\bibnamefont {Zamponi}},\ }\href@noop {}
  {\bibfield  {journal} {\bibinfo  {journal} {Phys. Rev. Lett.}\ }\textbf
  {\bibinfo {volume} {96}},\ \bibinfo {pages} {065702} (\bibinfo {year}
  {2006})}\BibitemShut {NoStop}%
\bibitem [{\citenamefont {Antenucci}\ \emph {et~al.}(2015)\citenamefont
  {Antenucci}, \citenamefont {Conti}, \citenamefont {Crisanti},\ and\
  \citenamefont {Leuzzi}}]{a63}%
  \BibitemOpen
  \bibfield  {author} {\bibinfo {author} {\bibfnamefont {F.}~\bibnamefont
  {Antenucci}}, \bibinfo {author} {\bibfnamefont {C.}~\bibnamefont {Conti}},
  \bibinfo {author} {\bibfnamefont {A.}~\bibnamefont {Crisanti}},\ and\
  \bibinfo {author} {\bibfnamefont {L.}~\bibnamefont {Leuzzi}},\ }\href@noop {}
  {\bibfield  {journal} {\bibinfo  {journal} {Phys. Rev. Lett.}\ }\textbf
  {\bibinfo {volume} {114}},\ \bibinfo {pages} {043901} (\bibinfo {year}
  {2015})}\BibitemShut {NoStop}%
\bibitem [{\citenamefont {Antenucci}\ \emph {et~al.}(2016)\citenamefont
  {Antenucci}, \citenamefont {Crisanti}, \citenamefont {Ibáñez-Berganza},
  \citenamefont {Marruzzo},\ and\ \citenamefont {Leuzzi}}]{rev1}%
  \BibitemOpen
  \bibfield  {author} {\bibinfo {author} {\bibfnamefont {F.}~\bibnamefont
  {Antenucci}}, \bibinfo {author} {\bibfnamefont {A.}~\bibnamefont {Crisanti}},
  \bibinfo {author} {\bibfnamefont {M.}~\bibnamefont {Ibáñez-Berganza}},
  \bibinfo {author} {\bibfnamefont {A.}~\bibnamefont {Marruzzo}},\ and\
  \bibinfo {author} {\bibfnamefont {L.}~\bibnamefont {Leuzzi}},\ }\href@noop {}
  {\bibfield  {journal} {\bibinfo  {journal} {Phil. Mag.}\ }\textbf {\bibinfo
  {volume} {96}},\ \bibinfo {pages} {704} (\bibinfo {year} {2016})}\BibitemShut
  {NoStop}%
\bibitem [{\citenamefont {Gradenigo}\ \emph {et~al.}(2020)\citenamefont
  {Gradenigo}, \citenamefont {Antenucci},\ and\ \citenamefont
  {Leuzzi}}]{rev2b}%
  \BibitemOpen
  \bibfield  {author} {\bibinfo {author} {\bibfnamefont {G.}~\bibnamefont
  {Gradenigo}}, \bibinfo {author} {\bibfnamefont {F.}~\bibnamefont
  {Antenucci}},\ and\ \bibinfo {author} {\bibfnamefont {L.}~\bibnamefont
  {Leuzzi}},\ }\href {https://doi.org/10.1103/PhysRevResearch.2.023399}
  {\bibfield  {journal} {\bibinfo  {journal} {Phys. Rev. Res.}\ }\textbf
  {\bibinfo {volume} {2}},\ \bibinfo {pages} {023399} (\bibinfo {year}
  {2020})}\BibitemShut {NoStop}%
\bibitem [{\citenamefont {Niedda}\ \emph {et~al.}(2023)\citenamefont {Niedda},
  \citenamefont {Gradenigo}, \citenamefont {Leuzzi},\ and\ \citenamefont
  {Parisi}}]{rev2}%
  \BibitemOpen
  \bibfield  {author} {\bibinfo {author} {\bibfnamefont {J.}~\bibnamefont
  {Niedda}}, \bibinfo {author} {\bibfnamefont {G.}~\bibnamefont {Gradenigo}},
  \bibinfo {author} {\bibfnamefont {L.}~\bibnamefont {Leuzzi}},\ and\ \bibinfo
  {author} {\bibfnamefont {G.}~\bibnamefont {Parisi}},\ }\href@noop {}
  {\bibfield  {journal} {\bibinfo  {journal} {SciPost Phys.}\ }\textbf
  {\bibinfo {volume} {14}},\ \bibinfo {pages} {144} (\bibinfo {year}
  {2023})}\BibitemShut {NoStop}%
\bibitem [{\citenamefont {Palacios}\ \emph {et~al.}(2023)\citenamefont
  {Palacios}, \citenamefont {Siqueira}, \citenamefont {de~Ara\'ujo},
  \citenamefont {Gomes},\ and\ \citenamefont {Raposo}}]{rev3}%
  \BibitemOpen
  \bibfield  {author} {\bibinfo {author} {\bibfnamefont {G.}~\bibnamefont
  {Palacios}}, \bibinfo {author} {\bibfnamefont {A.~C.~A.}\ \bibnamefont
  {Siqueira}}, \bibinfo {author} {\bibfnamefont {C.~B.}\ \bibnamefont
  {de~Ara\'ujo}}, \bibinfo {author} {\bibfnamefont {A.~S.~L.}\ \bibnamefont
  {Gomes}},\ and\ \bibinfo {author} {\bibfnamefont {E.~P.}\ \bibnamefont
  {Raposo}},\ }\href@noop {} {\bibfield  {journal} {\bibinfo  {journal} {Phys.
  Rev. A}\ }\textbf {\bibinfo {volume} {107}},\ \bibinfo {pages} {063510}
  (\bibinfo {year} {2023})}\BibitemShut {NoStop}%
\bibitem [{\citenamefont {Hackenbroich}\ \emph {et~al.}(2002)\citenamefont
  {Hackenbroich}, \citenamefont {Viviescas},\ and\ \citenamefont
  {Haake}}]{rev4}%
  \BibitemOpen
  \bibfield  {author} {\bibinfo {author} {\bibfnamefont {G.}~\bibnamefont
  {Hackenbroich}}, \bibinfo {author} {\bibfnamefont {C.}~\bibnamefont
  {Viviescas}},\ and\ \bibinfo {author} {\bibfnamefont {F.}~\bibnamefont
  {Haake}},\ }\href@noop {} {\bibfield  {journal} {\bibinfo  {journal} {Phys.
  Rev. Lett.}\ }\textbf {\bibinfo {volume} {89}},\ \bibinfo {pages} {083902}
  (\bibinfo {year} {2002})}\BibitemShut {NoStop}%
\bibitem [{\citenamefont {de~Ara\'ujo}\ \emph {et~al.}(2014)\citenamefont
  {de~Ara\'ujo}, \citenamefont {Roslund}, \citenamefont {Cai}, \citenamefont
  {Ferrini}, \citenamefont {Fabre},\ and\ \citenamefont {Treps}}]{ti_safira}%
  \BibitemOpen
  \bibfield  {author} {\bibinfo {author} {\bibfnamefont {R.~M.}\ \bibnamefont
  {de~Ara\'ujo}}, \bibinfo {author} {\bibfnamefont {J.}~\bibnamefont
  {Roslund}}, \bibinfo {author} {\bibfnamefont {Y.}~\bibnamefont {Cai}},
  \bibinfo {author} {\bibfnamefont {G.}~\bibnamefont {Ferrini}}, \bibinfo
  {author} {\bibfnamefont {C.}~\bibnamefont {Fabre}},\ and\ \bibinfo {author}
  {\bibfnamefont {N.}~\bibnamefont {Treps}},\ }\href@noop {} {\bibfield
  {journal} {\bibinfo  {journal} {Phys. Rev. A}\ }\textbf {\bibinfo {volume}
  {89}},\ \bibinfo {pages} {053828} (\bibinfo {year} {2014})}\BibitemShut
  {NoStop}%
\end{thebibliography}%

\end{document}


\renewcommand{\vec}{\mathbf}

\title{Supplemental Material for: Observation of replica symmetry breaking in standard mode-locked fiber laser}


\author{Nicolas P. Alves}
\affiliation{%
 Departamento de F\'{\i}sica, Universidade Federal de Pernambuco, 50670-901 Recife, PE, Brazil
}%
\author{Wesley F. Alves}
\affiliation{%
 Departamento de F\'{\i}sica, Universidade Federal de Pernambuco, 50670-901 Recife, PE, Brazil
}%
\author{Andre C. A. Siqueira}
\affiliation{%
 Departamento de F\'{\i}sica, Universidade Federal de Pernambuco, 50670-901 Recife, PE, Brazil
}%
\author{Naudson L. L. Matias}
\affiliation{%
 Departamento de F\'{\i}sica, Universidade Federal de Pernambuco, 50670-901 Recife, PE, Brazil
}%
\author{Anderson S. L. Gomes}
\affiliation{%
 Departamento de F\'{\i}sica, Universidade Federal de Pernambuco, 50670-901 Recife, PE, Brazil
}%
\author{Ernesto P. Raposo}
\affiliation{Laborat\'orio de F\'{\i}sica Te\'orica e Computacional, Departamento de F\'{\i}sica, Universidade Federal de Pernambuco, 50670-901 Recife, PE, Brazil}
\author{Marcio H. G. de Miranda}
\affiliation{%
 Departamento de F\'{\i}sica, Universidade Federal de Pernambuco, 50670--901 Recife, PE, Brazil
}%
\email{marcio.miranda@ufpe.br}


\maketitle

\subsection{Experimental setup for measurements in the MLFL system}

The experimental setup, shown in Fig.~\ref{MS1}, consists of a homemade ytterbium-based mode-locked fiber laser (MLFL) system supporting passive standard mode-locking (SML) lasing regime. 
The pump source is a CW fiber-coupled diode laser (LD) emitting at 976 nm, coupled into the gain medium through a wavelength division multiplexer (WDM) with an efficiency of 89$\%$. 
We go backwards in the pump laser calibration (current from 800~mA to~zero) since it is easier to reach the SML regime, due to the system's hysteretic behavior~\cite{Lucas18, Campos21}.
The gain medium is a highly Yb-doped fiber CorActive Yb214 (YDF), with an absorption coefficient of 1348 dB/m at 976 nm and length of 22 cm~\cite{Lucas18, Cecilia19}. 

A pair of GRIN collimators (A1) are used to couple the beam from the fiber medium to the free-space section and then back to the fiber medium, providing 35$\%$ of optical coupling efficiency. 
A unidirectional operation is guaranteed by an optical isolator (OI). The laser cavity includes a grating pair (A5) with 600 lines/mm and reflectivity of 65$\%$ per pass, used to manage the total intracavity dispersion. 
The distance between the grating pair sets the operation of the laser at either normal or anomalous net dispersion regime. 
Here the grating pair is separated by 3.0~cm, leading to a slightly anomalous overall cavity dispersion. 

The laser achieves passive mode-locking via nonlinear polarization rotation (NPR)~\cite{Ferman}, which involves a set of waveplates comprising two quarter-wave (A2) and one half-wave (A3) plates, and a polarizing beam splitter (PBS) that controls polarization and also serves as output coupler. 
The MLFL system produces optical spectra centered around 1025~nm and a fundamental repetition rate of 120~MHz in the SML regime. 

Data acquisition is accomplished by directing the output beam from the cavity through a beamsplitter (A7), with a portion sent to an optical spectrum analyzer (Ocean Optics HR4000, resolution 0.24~nm) (B5) and the remaining portion sent to a photodetector (1~GHz bandwidth) (A8).
The signal from the photodetector is directed toward a RF spectrum analyzer (Keysight N9340B) (B3) and a 1 GHz oscilloscope (Keysight DSO7104B) (B4) for regime operation identification. 

For replica symmetry breaking (RSB) investigation, it is essential to guarantee the same experimental conditions from replica to replica. For this reason, the data acquisition is automated using a custom LabVIEW and Python programs executed on a laptop (B6). 
The LabVIEW program controls the temperature (constant at 25$^{o}$C) and electric current of the diode laser, adjustable in 1.2~mA steps. 
The Python script collects 3000 optical spectra (replicas), for each current setting. 
In the results presented in the main text, each replica has an integration time of 9~ms and there is a time interval of 130~ms between consecutive replicas. 
In Section C below, we probe the robustness of our findings with respect to different integration times.

\setcounter{figure}{0}
\renewcommand{\thefigure}{S\arabic{figure}}

\begin{figure}[t] 
\includegraphics[width=0.51\textwidth]{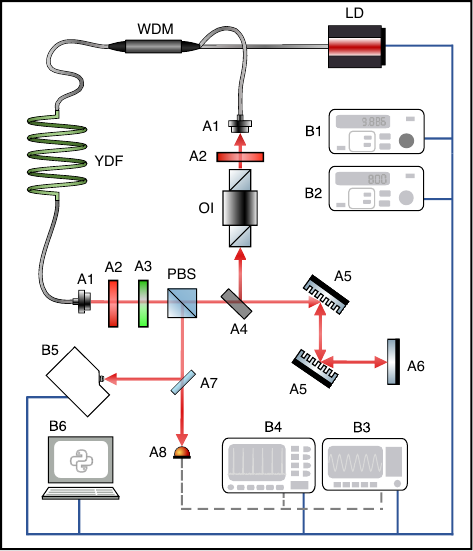}
\caption{Experimental setup of the Yb-based mode-locked fiber laser (MLFL): current (B1) and temperature (B2) controllers, pump diode laser (LD), wavelength division multiplexer (WDM), gain medium Yb-doped fiber (YDF), pair of GRIN collimators (A1), quarter-wave plate (A2), half-wave plate (A3), grating pair (A5), mirrors (A4, A6), beam spliter (A7), optical isolator (OI), and polarizing beam splitter (PBS). Data acquisition system: photodetector (A8), RF spectrum analyzer (B3), oscilloscope (B4), optical spectrum analyzer (B5), and laptop (B6).}
\label{MS1}
\end{figure}

Prior to the automated data collection, the current is manually reduced to determine the value for which the mode-locked regime is lost.
An average threshold point is obtained and the automation is set to begin a few mA above this point, in which mode-locked operation is still present. 
The automation is set to collect data around 100 current points, corresponding to a range of currents from 410 mA to 200 mA.
%


\subsection{Spectra, temporal trace, and average SML pulse  duration in the MLFL system}

\begin{figure}[t] 
\includegraphics[width=0.6\textwidth]{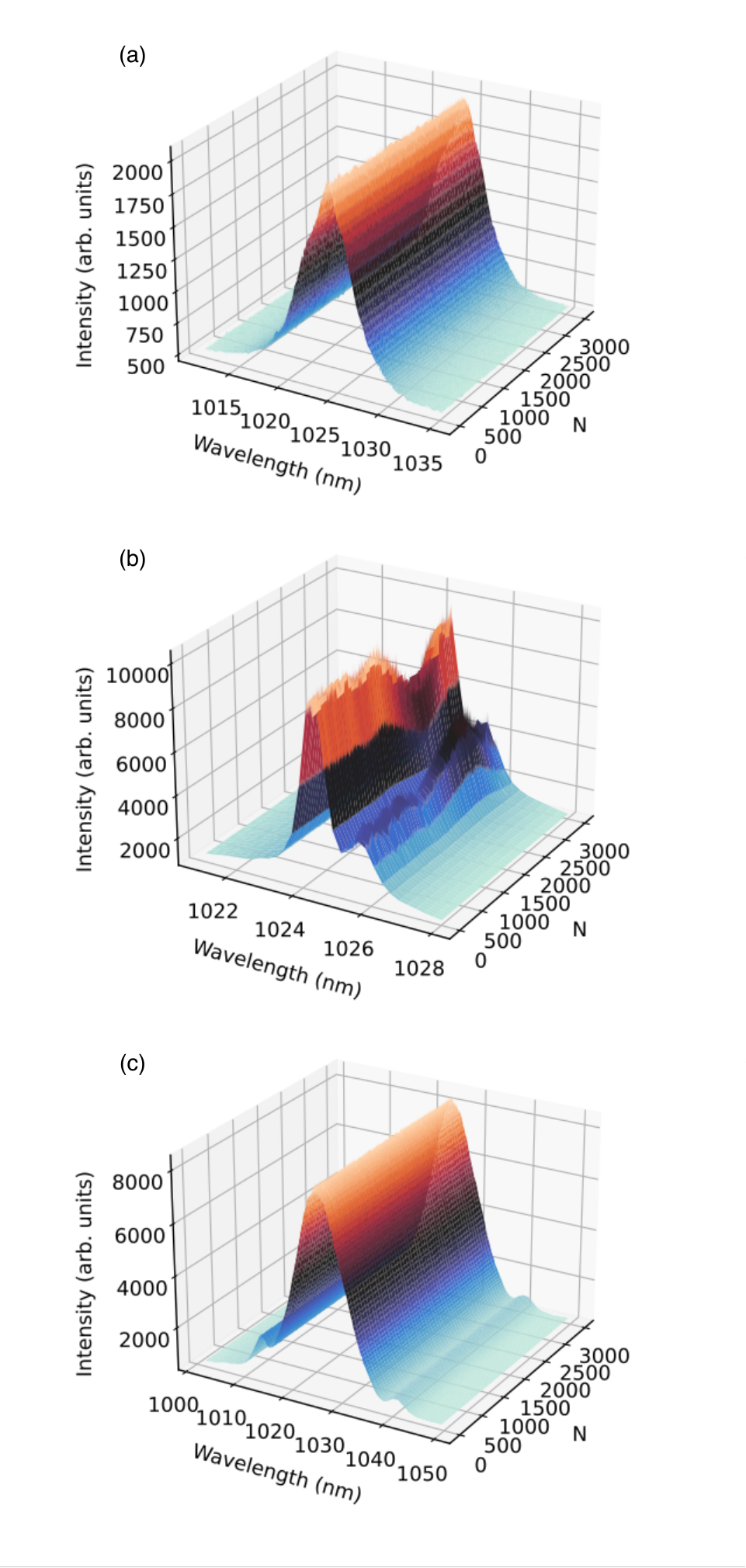}
\caption{3000 spectra for each of the three relevant regimes in the MLFL system: (a) CW (excitation current 274~mA), (b) QML (392~mA, near the transition to the SML phase), and (c)~SML (407~mA).}
\label{SM1B}
\end{figure}

We display in Fig.~\ref{SM1B} the whole temporal sequence of 3000 recorded spectra for each of the three relevant regimes of the MLFL system: 
(a) CW (excitation current 274~mA), 
\linebreak 
(b)~QML (392~mA, near the transition to the SML phase), and (c) SML (407~mA) (these current values also correspond to those in Fig.~2 of the main text). 
As expected, the intermediate QML regime presents a more rugged spectral pattern with somewhat larger fluctuations, if compared to the much smoother, weakly fluctuating profiles of the CW and SML phases.

\begin{figure}[t] 
\includegraphics[width=0.55\textwidth]{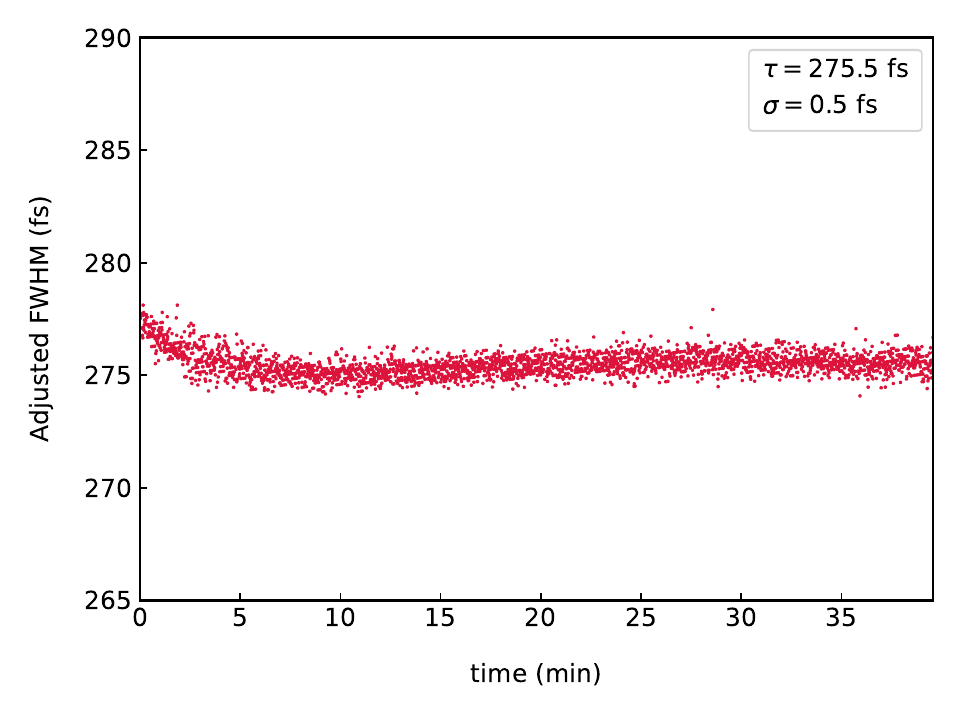}
\caption{FWHM as a function of time calculated from all 3000 autocorrelation traces in the SML regime of the MLFL system. The average temporal duration of the typical pulse trains in this regime is $\tau = 275.5$~fs, with standard deviation of 0.55~fs.}
\label{SM1C}
\end{figure}

In addition, from the FWHM measurement we have also inferred the temporal duration~$\tau$ of the typical pulse trains in the SML regime. 
Figure~\ref{SM1C} shows the FWHM as a function of time calculated from all 3000 autocorrelation traces. 
The average value calculated from the data is $\tau = 275.5$~fs, with standard deviation of 0.55~fs. 
The curve does not show any kind of (anti)correlation. 
A small linear drift of slope $-0.26$~fs/min has been filtered in the statistical analysis of Fig.~\ref{SM1C}.

\begin{figure}[t] 
\includegraphics[width=0.94\textwidth]{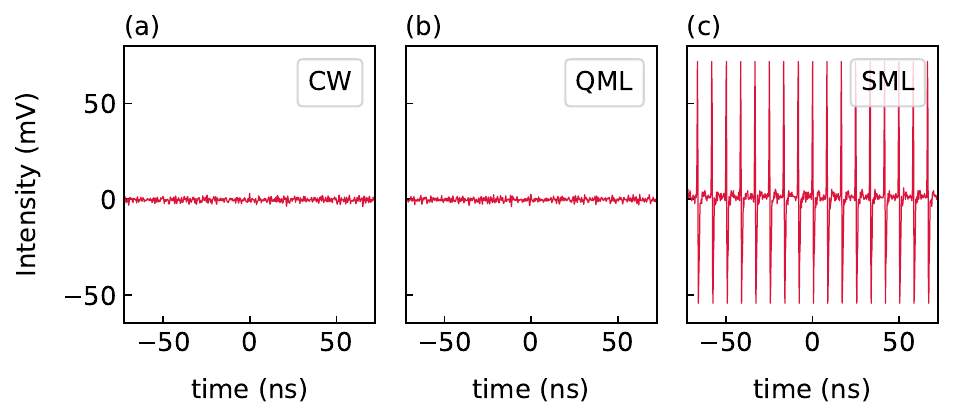}
\caption{Temporal traces of the (a) CW (106 mA), (b) QML (321 mA), and (c) and SML (405~mA) regimes. Light is emitted in the form of ultrashort pulses with typical average duration of around 275~fs in the SML regime, consistent with Fig.~\ref{SM1C}.}
\label{SM4new}
\end{figure}

In Fig.~\ref{SM4new} we show the temporal traces of the CW, QML, and SML regimes. 
A clear distinction between these phases can be noticed. 
In particular, in the SML regime we observe that light is emitted in the form of ultrashort pulses with typical average duration of around 275~fs, consistent with Fig.~\ref{SM1C}. 
Such pulses are absent both in the CW and~QML~regimes.

In contrast, while the $P(q)$ profile is similar in the replica-symmetric CW and SML phases, a distinguished bimodal $P(q)$ distribution emerges in the RSB QML regime. 
So the analysis of both the temporal trace and Parisi distribution helps setting the photonic~regime. 
%




\subsection{Results for different integration times}

\begin{figure} 
\includegraphics[width=0.8\textwidth]{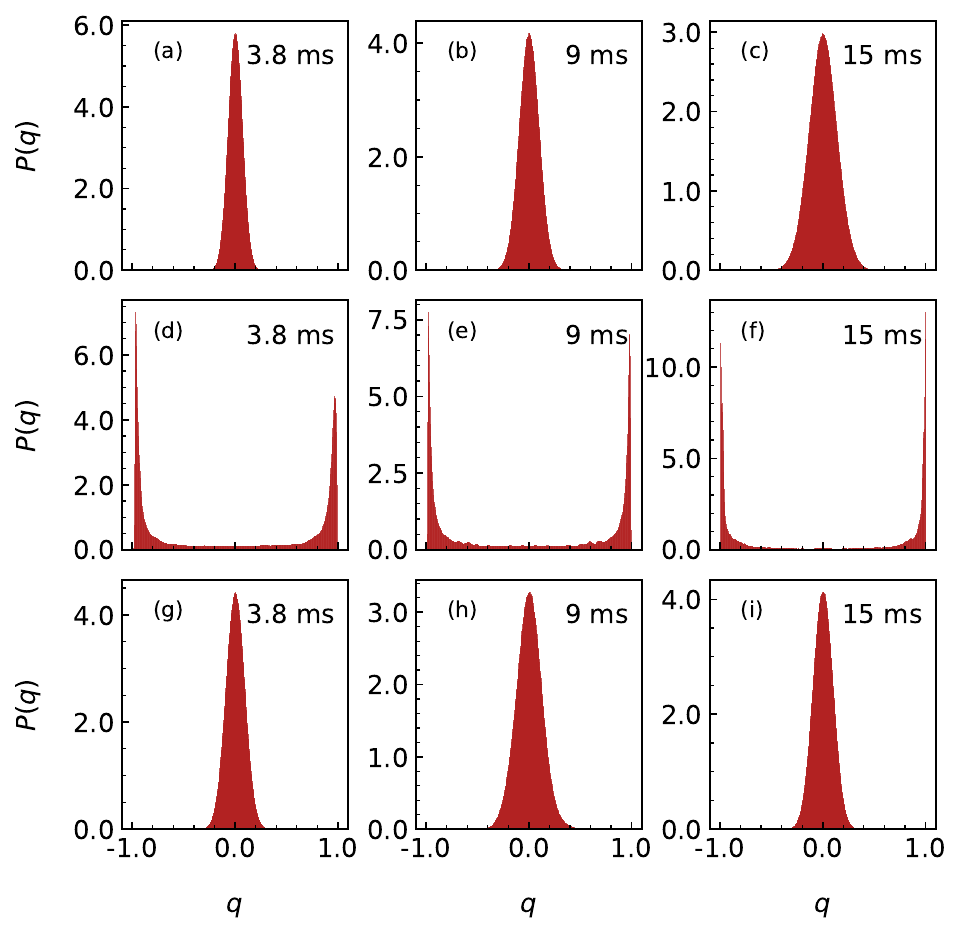}
\caption{Distribution $P(q)$ of Parisi overlap parameter for different integration times: 3.8~ms, 9~ms (main text for comparison), and 15~ms. Top and bottom lines: replica-symmetric (a)-(c) CW (106~mA) and (g)-(i) SML (405~mA) regimes. Middle line: (d)-(f) RSB QML regime (321~mA).}
\label{SM5new}
\end{figure}

In order to probe whether our findings are robust concerning the integration time used to acquire the optical spectra, we show in Fig.~\ref{SM5new} the results for $P(q)$ using acquisition times of 3.8~ms, 9~ms (as in the main text, for comparison), and 15~ms.

The first line in Fig.~\ref{SM5new} corresponds to the CW regime (excitation current~106 mA), with a replica-symmetric profile of $P(q)$ centered around $q = 0$. 
The middle line shows the RSB QML regime (321~mA), with two side maxima of $P(q)$ near $q = \pm 1$.
Finally, the replica-symmetric SML phase (405~mA) is depicted at the bottom line. 
%

\newpage


$ $

\newpage


These plots point to the robustness of our findings concerning different integration times.



\subsection{Theoretical modeling of photonic RSB regimes}

In the last two decades, the photonics research group of Universit\`a di Roma ``Sapienza" has remarkably extended~\cite{a61,a63} the concept of RSB from disordered magnetic systems to nonlinear optical systems with some sort of disorder.
Here we briefly review the theoretical~modeling underlying the photonic RSB regimes. 
For a nice review, see~\cite{rev1}  (see~also~\cite{rev2b,rev2,rev3}).

The Hamiltonian formulation of nonlinear multimode photonic systems takes as starting point the system-and-bath Hamiltonian model of quantum nonlinear photonic systems~\cite{rev4}.
In the semiclassical context, the creation and annihilation operators of the electromagnetic field are replaced by their expected values, yielding the effective Hamiltonian function, 
\begin{equation} 
\label{eq:1}
{\cal H} = \sum_{\{k_1, k_2\}'} g_{k_1,k_2}^{(2)}a_{k_1}a_{k_2}^* + \sum_{\{k_1, k_2,k_3,k_4\}'} g_{k_1,k_2,k_3,k_4}^{(4)}a_{k_1}a_{k_2}^*a_{k_3}a_{k_4}^*,
\end{equation}
where $a_k$ denotes the mode amplitudes in the slow amplitude approximation, and $g_{k_1,k_2}^{(2)}$ and $g_{k_1,k_2,k_3,k_4}^{(4)}$ are, respectively, quadratic and quartic mode couplings, with the latter associated with $\chi^{(3)}$-susceptibility.  
In disorder-free laser systems one~has $\mbox{Re} \{ g^{(2)}_{kk} \} = \alpha_k - \gamma_k$, with $\gamma_k$ and $\alpha_k$ as the gain and radiation loss coefficients, respectively, while the quartic coupling is related to nonlinearities in the refractive index. 
For systems with some degree of disorder in the strong cavity limit of small losses, the diagonal elements of $g_{k_1,k_2}^{(2)}$ dominate over off-diagonal ones and $g_{k_1,k_2,k_3,k_4}^{(4)}$ carries the signature of the randomness of the active~medium.

In the slow amplitude modes, with $a_k (t) = A_k (t) e^{i \phi_k (t)}$, the dynamics of the amplitudes~$A_k$ evolves much slower than that of the phases~$\phi_k$. 
Upon Fourier transform one has ${\bar{a}_k(\omega) \approx \delta (\omega - \omega_k)}$, where $\delta$~is~the Dirac delta function. 
This implies that only combinations of mode frequencies such as ${| \omega_{k_1} - \omega_{k_2} | \lesssim \gamma}$ and ${| \omega_{k_1} - \omega_{k_2} + \omega_{k_3} - \omega_{k_4} | \lesssim \gamma}$ are accounted for in the first and second terms of Eq.~(\ref{eq:1}), respectively, where $\gamma$~is~the~modes~linewidth.
Such constraints yield frequency matching conditions indicated by $\{k_1, k_2\}'$~and $\{k_1, k_2,k_3,k_4\}'$ in Eq.~(\ref{eq:1}).
The slow amplitude modes thus form a suitable basis to describe lasing~modes.

The calculation from first principles of the quadratic and quartic couplings in Eq.~(\ref{eq:1}) is virtually unfeasible. 
One possibility is to assume~\cite{a61,a63,rev1} coupling values taken from Gaussian distributions, with mean and variance respectively related to the excitation energy and disorder strength in a mean-field-type approach. 
In a statistical physics calculation using the replica trick~\cite{a61,a63,rev1}, a rich phase diagram has been theoretically proposed as a function of the excitation energy and disorder strength, with the former corresponding to the inverse temperature and modes playing the role of spins in the magnetic-to-photonic~analogy.

In this analogy, the high-temperature paramagnetic phase corresponds to the CW regime for excitation energies below the lasing threshold and any disorder strength. 
In this regime modes are uncorrelated and oscillate independently, with light emission in the form of continuous waves. 
In contrast, the SML regime above threshold displays ultrashort pulse emission, either in a disorder-free replica-symmetric state with coherent oscillation of all modes (analogous to the ferromagnetic phase) or in a RSB regime with the distributions of mode couplings presenting not too large variances and most modes oscillating coherently (random bond ferromagnet). 
Moreover, the photonic counterpart of the magnetic spin glass phase corresponds to the photonic RSB glassy behavior above threshold, demonstrated, e.g., in random laser and random fiber laser systems.
In this case modes competition for gain is enhanced with larger variances of the distributions of mode couplings, leading to frustrated synchronous oscillation of nontrivially correlated modes captured by Parisi overlap parameter~$q$. 
In the present glassy QML regime with RSB, emission does~not occur in the form of ultrashort pulses (see Fig.~\ref{SM4new}), phase coherence is not achieved, and the multiple longitudinal modes take part in a randomly frustrated competition for gain with spin-glass-type \linebreak correlations depicted in the bimodality of the  distribution $P(q)$, shown in Figs.~2(b)-(c) of the main text and Fig.~\ref{SM5new}. 
Lastly, we should also mention a phase locking wave regime in the theoretical phase diagram which has no magnetic analogue~\cite{a63,rev1}.
%


On the numerical side, the works~\cite{rev2b,rev2} have applied a parallel tempering (exchange) Monte Carlo algorithm based on a mode-locked four-phasor Hamiltonian model for random laser systems to investigate the probability distribution $P(q)$ of the Parisi overlap parameter~$q$, among other observables. 
Although the local side maxima of $P(q)$ around $q \approx \pm 1$ have their onset at the random lasing threshold, they are not global maxima. Instead, the replica-symmetric global maximum of $P(q)$ at $q = 0$ still prevails even above threshold.

More recently, a distinct Monte Carlo approach~\cite{rev3} employed photonic random walkers that diffuse and get randomly scattered in the active medium, interacting not only with the population of excited atoms but also among themselves in a mean-field-type approach. 
In this case, $P(q)$ displays global maxima around $q \approx \pm 1$ above threshold for proper choices of the mean and variance of the Gaussian interactions between modes and probability of the photonic walkers of getting randomly scattered, among other parameters.
%


\vspace*{1 cm} 

\subsection{Replica symmetry in the SML regime of titanium-sapphire mode-locked~laser}

We also employed a titanium-sapphire mode-locked laser~\cite{ti_safira} with the aim to investigate if no RSB is present in the SML regime, just like for the Yb-based  MLFL in the same~regime.

\begin{figure}[t] 
\includegraphics[width=0.8\textwidth]{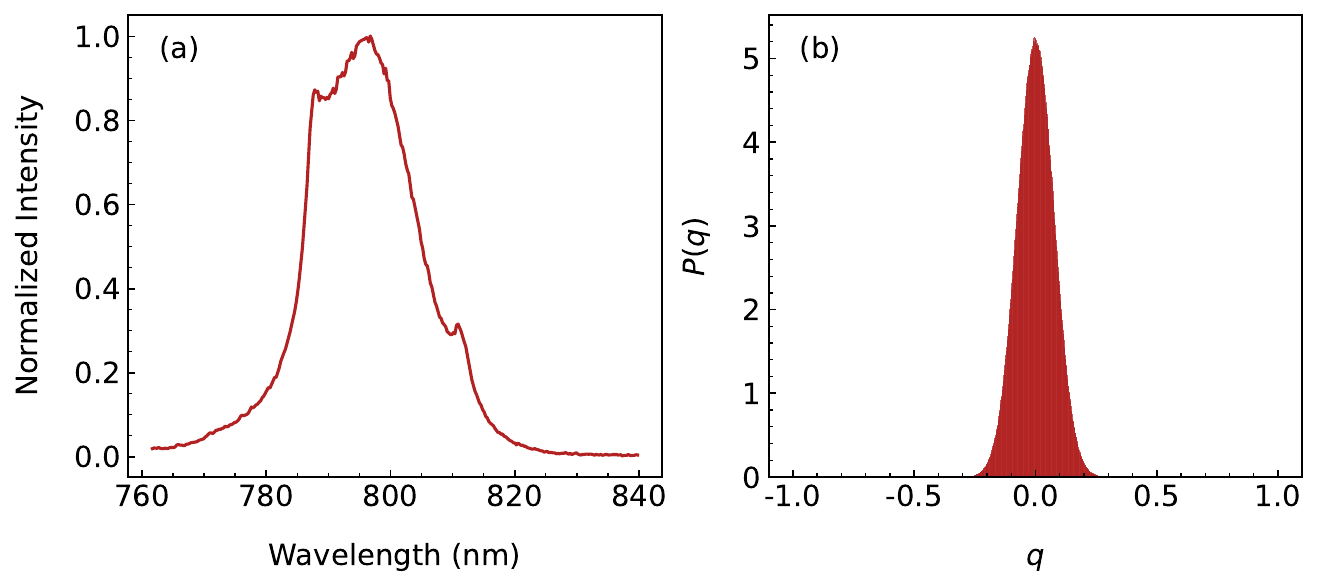}
\caption{Mode-locked titanium-sapphire laser. (a) Typical spectrum in the SML regime for the pump laser power of 4.7 W. (b) Distribution $P(q)$ of Parisi overlap parameter~$q$ in the SML regime displaying the pronounced central maximum at $q = 0$, typical of the replica symmetric phase.}
\label{SM2}
\end{figure}

We have found that the Ti:Sa laser displays replica symmetric behavior for the interval of excitation currents investigated in the SML regime. 
Indeed, as shown in Fig.~\ref{SM2} in this regime for the pump laser power of 4.7~W, the distribution $P(q)$ of Parisi overlap parameter~$q$ presents the pronounced central maximum at $q = 0$, typical of the replica symmetric~phase.

We have not gone through the whole phase diagram of the Ti:Sa laser to discard a QML-like regime with RSB in this system. 
In this sense, a more thorough investigation on the photonic regimes of the Ti:Sa laser for a broader range of currents is left to a future work.
%


\subsection{Replica symmetry in the CW pump sources of MLFL and Ti:sapphire lasers}

Both Yb-based MLFL and titanium-sapphire systems are pumped by CW sources. 
In an investigation of the emergence of the RSB phenomenon in any photonic system, it is essential to discard that the RSB behaviour does not arise as a consequence of the pump source properties.

\begin{figure}[t] 
\includegraphics[width=0.8\textwidth]{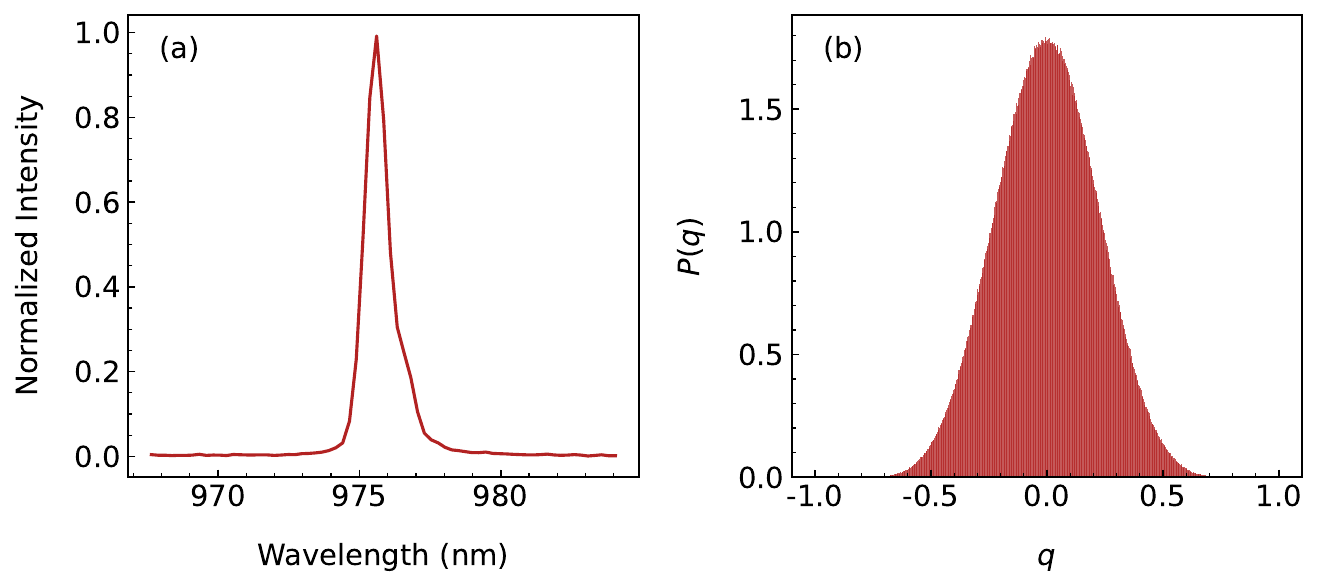}
\caption{CW pump laser of the Yb-based MLFL system. (a) Typical CW spectrum for 321.6~mA. (b) Distribution $P(q)$ of Parisi overlap parameter with the maximum at $q = 0$ indicating replica symmetric behavior.}
\label{SM3}
\end{figure}

\begin{figure}[t] 
\includegraphics[width=0.8\textwidth]{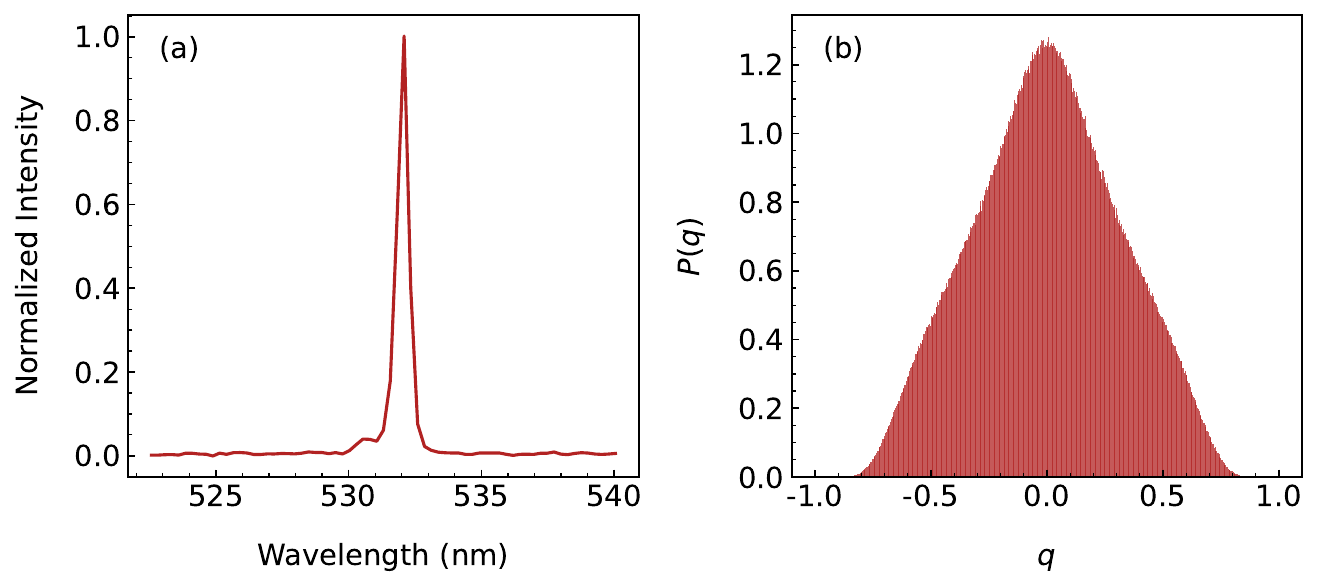}
\caption{The same as in Fig.~\ref{SM3} but for the CW pump source of the Ti:sapphire laser at 4.7 W.}
\label{SM4}
\end{figure}

We thus additionally analyzed the distribution $P(q)$ of the Parisi parameter for a wide range of excitation currents of the CW pump sources, especially for the current values that yield RSB behaviour in the QML regime of the Yb-based MLFL. 
Results displayed in Figs.~\ref{SM3} and~\ref{SM4}, respectively for the pump sources of the MLFL and Ti:sapphire systems, leave no doubt about the replica symmetric emission regime of the pump lasers. 
We, therefore, conclude that the RSB features of the Yb-based MLFL system cannot be attributed to the pump source properties. 

\vspace*{0.5 cm}

\bibliography{supplemental}